\documentclass[10pt,aps,a4paper,prd,floatfix,preprintnumbers,twocolumn,notitlepage,nofootinbib]{revtex4-1}
\usepackage[utf8]{inputenc}
\usepackage[british,UKenglish,USenglish,american]{babel}
\usepackage{amsmath}
\usepackage{graphicx}
\usepackage{subcaption}
\usepackage{mathrsfs}
\usepackage{amssymb}
\usepackage{bbm}
\usepackage{nicefrac}
\usepackage{microtype}
\usepackage{physics}
\usepackage{xcolor}
\usepackage{pifont}
\usepackage{times}
\usepackage{slashed}
\usepackage{mathtools}
\usepackage{epsfig}
\usepackage{textcomp}
\usepackage{threeparttable}
\usepackage{multirow}
\newcommand{\p}{\prime}
\newcommand{\lag}{{\cal L}}
\newcommand{\e}{Eq.}

\renewcommand{\bar}{\overline}
\newcommand{\be}{\begin{equation}}
\newcommand{\ee}{\end{equation}}
\newcommand{\ba}{\begin{eqnarray}}
\newcommand{\ea}{\end{eqnarray}}
\newcommand{\f}{Fig.}
\newcommand{\fs}{Figs.}
\newcommand{\amu}{$\Delta a_{\mu}$}
\newcommand{\am}{$a_\mu$}
\newcommand{\gmu}{$g_\mu -2$ }
\newcommand{\su}{$SU(4)_L\otimes U(1)_X$ }

\begin{document} 
\title{Are 3-4-1 models able to explain the upcoming results of the muon anomalous magnetic moment?}
\author{D. Cogollo$^1$}
\email{diegocogollo@df.ufcg.edu.br}
\author{Yohan M. Oviedo-Torres$^{1,4}$}
\email{ymot@estudantes.ufpb.br}
\author{Yoxara S. Villamizar$^{2,3}$}
\email{yoxara@ufrn.edu.br}
\affiliation{$^1$Departamento de F\'isica, Universidade Federal de Campina Grande, Caixa Postal 10071, 58109-970, Campina Grande, PB, Brazil \\
$^2$International Institute of Physics, Universidade Federal do Rio Grande do Norte, Campus Universit\'ario, Lagoa Nova, Natal-RN 59078-970, Brazil\\
$^3$Departamento de F\'isica Teórica e Experimental, Universidade Federal do Rio Grande do Norte, Natal - RN, 59078-970, Brazil\\
$^4$Departamento de F\'isica, Universidade Federal da Para\'iba,
Caixa Postal 5008, 58051-970, Jo\~ao Pessoa, PB, Brazil\\
}

\begin{abstract}
In light of the upcoming measurement of the muon anomalous magnetic moment (g-2), we revisit the corrections to g-2 in the context of the $SU(4)_L \times U(1)_X$ gauge symmetry. We investigate three models based on this gauge symmetry and express our results in terms of the energy scale at which the  $SU(4)_L \times U(1)_X$ symmetry is broken. To draw solid conclusions we put our findings into perspective with existing collider bounds. Lastly, we highlight the difference between our results and those rising from $SU(3)_L \times U(1)_X$ constructions.

\end{abstract}

\maketitle
\flushbottom

\section{Introduction}
\label{intro}
\begin{figure*}[!htp] 
    \centering
    \includegraphics[scale=0.5]{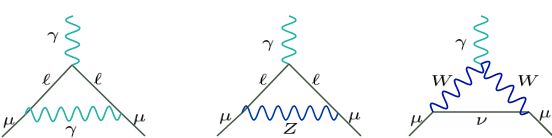}
    \caption{Feyman diagrams of the corrections to \am on SM electroweak interactions.}
    \label{feynmandiagram}
\end{figure*}

Currently, the Standard Model (SM) of particle physics needs to be extended to explain signals or explore evidences of new physics like dark matter, neutrino masses, flavor universality violation, etc. There are different ways to extend the SM, and
these ways open several alternatives to do physics beyond the SM. For example, to extend the SM gauge symmetry implies the existence of new gauge bosons. At least, by extending the symmetry just by a $U(1)_{x}$ group, we are predicting the existence of a new $Z$ boson. By extending the $SU(2)_{L}$ symmetry to a larger $SU(N)_{L}$ group, we have $N^{2}-4$ new gauge bosons at our disposal. Models based on the $SU(3)_{C} \times SU(4)_{L} \times U(1)_{X}$ 3-4-1 symmetry \cite{Pisano:1994tf}, are that kind of beyond SM models. In this work we will focus in a very fundamental problem of particle physics, the anomalous magnetic moment of the muon, that we will describe in some detail below. The idea here is to study the different  contributions to that anomaly arising in different versions of the 3-4-1 model. Our work is based on a rigorous correction to a numerical analysis previously carried out \cite{Cogollo:2014tra,Cogollo:2014aka,Cogollo:2015fpa}, taking into account the most updated, model independent, analytical expressions that contributes to the anomaly \cite{Lindner:2016bgg}, when compared to previous works \cite{Queiroz:2014zfa} . It is important to mention that for elementary particles of mass $\textbf{m}$, electric charge $\textbf{q}$, and spin $\textbf{S}=\frac{1}{2}$, the Dirac equation predicts its magnetic dipole moment $\vec{\mu}$ , that it is an intrinsic property of the particle, given for the following relation:

\be
\overrightarrow{\mu} = g\frac{q}{2m}\overrightarrow{s},
\label{mu}
\ee

being $g$ the Landé g-factor or gyromagnetic ratio. For the muon, the prediction of the Dirac equation is $g_{\mu}=2$. Loop-level corrections generate little deviations from 2 -- the anomalous magnetic moment, parametrized by $a_{\mu} = (g_{\mu}-2)/2$. $a_{\mu}$ allows us to test the SM since each sector yields a sizeable correction \cite{Tanabashi:2018oca}, that represents interactions of the type $\mu \rightarrow \gamma \mu$, which can be seen in the Feynman diagrams of the \f \ref{feynmandiagram}. Nevertheless, there is a discrepancy between the Standard Model prediction and the experimental  measurements,  quantified by $\Delta a_{\mu}=a_{\mu}^{exp} - a_{\mu}^{SM}$, suggesting the presence of new physics that accounts for it. According  to  the  Particle  Data  Group  (PDG),  the  current  discrepancy reads $a_\mu= a_\mu^{exp}-a_\mu^{SM} = (261 \pm 78)\times 10^{-11}\,\, (3.3\sigma)$. The PDG review already acknowledges  recent  studies  where  the  significance  approaches $4\sigma$. However the large theoretical uncertainties can overshadow the significance of this discrepancy. It is important to mention that there are two experiments, the g-2 at FERMILAB \cite{Grange:2015fou} and the Muon g-2 at J-PARC \cite{Abe:2019thb}, that will be able to decrease the error bar and increase the discrepancy if the central value remains the same. Having in mind the g-2 collaboration is about to announce new results, we find important to review previous studies in this matter in the context of the 3-4-1 gauge symmetry. Along with the actual discrepancy reported by PDG of $(3.3\sigma)$, we will use the projected discrepancy of the $g-2$ collaboration, $\Delta a_\mu= a_\mu^{exp}-a_\mu^{SM} = (261 \pm 34)\times 10^{-11}\,\, (5\sigma)$, to impose the most stringent constraints on the scale of symmetry breaking and masses for three different versions of the 3-4-1 model, as aforementioned. The 3-4-1 symmetry is a natural extension of the $SU(3)_C \times SU(3)_L \times U(1)_X$ (3-3-1) symmetry that has been widely explored in the literature \cite{Pisano:1991ee,Foot:1992rh}. These 3-3-1 models can accommodate dark matter \cite{Fregolente:2002nx,Hoang:2003vj,deS.Pires:2007gi,Mizukoshi:2010ky,Profumo:2013sca,Dong:2013ioa,Dong:2013wca,Queiroz:2013lca,Kelso:2013nwa,Cogollo:2014jia,Dong:2014wsa,Dong:2014esa,Mambrini:2015sia,Alves:2016fqe,Carvajal:2017gjj,Dong:2017zxo,Montero:2017yvy,Arcadi:2017xbo,Huong:2019vej}, neutrino masses \cite{Montero:2000rh, Tully:2000kk,Montero:2001ts,Cortez:2005cp,Cogollo:2009yi,Queiroz:2010rj,Cogollo:2010jw,Cogollo:2008zc,Okada:2015bxa,Vien:2018otl,carcamoHernandez:2018iel,Nguyen:2018rlb,Pires:2018kaj,CarcamoHernandez:2019iwh,CarcamoHernandez:2019vih,CarcamoHernandez:2020pnh}, and also are entitled to a rich phenomenology concerning lepton flavor violation and collider physics \cite{Queiroz:2010rj,Alves:2011kc,Cogollo:2012ek,Alvares:2012qv,Alves:2012yp,Caetano:2013nya,Dong:2014wsa,Queiroz:2016gif,Ferreira:2019qpf}. 3-4-1 models embed these 3-3-1 models and therefore, we naturally inherit these features. As far as the muon anomalous magnetic moment is concerned several studies have been carried out in the past \cite{Kelso:2013zfa,Kelso:2014qka,Binh:2015cba,Binh:2015jfz,DeConto:2016ith,Cogollo:2017foz,Santos:2018qdx,deJesus:2020ngn}, but 3-4-1 models experience different contributions to g-2, and that motive us to explore them in perspective with existing bounds.  

In summary, we will investigate the corrections to \gmu in the context of 3-4-1 models. Our work is structured as follows: in Section \ref{models} we present the models; in Section \ref{results} we present our results; and later we conclude.

\section{\su Models}
\label{models}
  
The 3-4-1 model is an electroweak extension of the SM, which is based on $SU(3)_C \otimes SU(4)_L \otimes U(1)_X$ gauge symmetry. In general, 3-4-1 models were proposed to provide an elegant solution to the neutrinos masses, by placing the leptons $\nu,e,\nu^c$ and $e^c$ in the same multiplet of a $SU(4)_L$ \cite{Pisano:1994tf}. Today, we have different versions of the 3-4-1 model \cite{Dias:2013kma,Palcu:2009ks,Palcu:2009ky,Palcu:2009kb,Riazuddin:2008yx} each of them inherits the features of their respective 3-3-1 model \cite{Pisano:1991ee}. The most general expression for the electric charge operator in the case of the $SU(4)_L\otimes U(1)_X$ symmetry is given by:

\be
Q=aT_{3L}+\frac{b}{\sqrt{3}}T_{8L}+
\frac{c}{\sqrt{6}}T_{15L}+ XI_4,
    \label{Q}
\ee
where $a,b$ and $c$ are free parameters that allow us to set the fermion and scalar multiplets as well as the the gauge boson content. The $T_{iL}$ matrices are  the  generators of the $SU(4)_L$ group, defined as $T_{iL}=\lambda_{iL}/2$, being $\lambda_{iL}$ the Gell-Mann matrices for $SU(4)_L$. These gerenerators are normalized as Tr$(T_iT_j)=\delta_{ij}/2$. Also, in the \e  (\ref{Q}), $I_4$ is the $4\times 4$ identity  matrix and $X$ is a quantum number, equivalent to the hypercharge in the SM. In the next section, we will  briefly review the key theoretical aspects, which are relevant for the muon magnetic moment, for each one of the three different versions of the 3-4-1 model that we study here. Our goal is to reassess whether these models are capable of addressing the actual and projected discrepancy. In this way, we take interest in the interactions that can be represented as the Feynman diagram (\f \ref{feynmandiagram}), but instead of SM leptons $\ell$ and gauge bosons Z and W, new fermions and new gauge bosons called $Z^{\p}$, $W^{\p}$ and U, will mediate these interactions, as will be explained below.

\subsection{\su model with doubly charged gauge boson}
\label{model1}

In this model, the electric charge operator is defined as:

\begin{equation}
Q=\frac{1}{2}(\lambda_3-\frac{1}{\sqrt3}\lambda_8-\frac{2}{3}{\sqrt6}
\lambda_{15})+X,
\label{q}
\end{equation}

It is important to mention that in order to avoid anomalies, we must have the same number of $4$ and $4^*$ multiplets. For leptons, we have left and right-handed charged leptons and neutrinos in the same $SU(4)_L$ multiplet, that transform as $(1,4,0)$. The quark sector consists of one generation transforming as $(3,4,+2/3)$, and the two others as $(3,4^*,-1/3)$ \cite{Pisano:1994tf}. Concerning the right-handed quarks, they are all singlets under the symmetry in question. So, the fermionic content, excluding the right-handed quarks is:

\begin{align}
\begin{array}{ccccc}
f_{a L}=
\left(\begin{array}{c}
\nu_{a}\\
\ell_{a}\\
\nu_{a}^c\\
\ell_{a}^{c}\end{array}\right)_{L}\sim(1,4,0), \\  \\
Q_{1L}=
\left(\begin{array}{c}
u_{1}\\
d_{1}\\
u^\prime\\
J
\end{array}\right)_{L}\sim(3,4,2/3), \\ \\
Q_{\alpha L}=
\left(\begin{array}{c}
j_{\alpha}\\
d_{\alpha}^{\p}\\
u_\alpha\\
d_\alpha
\end{array}\right)_{L} \sim(3,4^*,-1/3).
\end{array}
\end{align}

where $a=1,2,3$ is a flavor index, counting the number of fermion families, and $\alpha=2,3$.

An interesting characteristic of this model is the presence of new fermions beyond the SM ones, they are two new quarks $u^{\prime}$ and $J$ with charges $+2/3$ and $+5/3$ respectively, and another four $j_{2,3}$ and $d_{2,3}^{\prime}$ with charges $-4/3$ and $-1/3$, respectively.

In order to generate masses for all the quarks it is necessary to  introduce three scalar multiplets $\eta$, $\rho$ and $\chi$, with just three of their neutral fields developing a vacuum expectation value, as we shown below:

\begin{align}
\begin{array}{ccccc}
\eta=
\left(\begin{array}{c}
\eta_{1}^0\\
\eta_{1}^{-}\\
\eta_2^0\\
\eta_2^+
\end{array}\right)
\sim(1,4,0), &
\left<\eta\right>=(v_1/\sqrt{2},0,0,0), \\ \\
\rho=
\left(\begin{array}{c}
\rho_{1}^+\\
\rho^{0}\\
\rho_2^+\\
\rho^{++}
\end{array}\right)
\sim(1,4,+1), &
\left<\rho\right>=(0,u/\sqrt{2},0,0), \\ \\
\chi=\left(\begin{array}{c}
\chi_{1}^-\\
\chi^{--}\\
\chi_2^-\\
\chi^{0}
\end{array}\right)
\sim(1,4,-1),&
\left<\chi\right>=(0,0,0,V_\chi/\sqrt{2}).
\end{array}
\end{align}

As for the charged leptons and neutrinos masses it is necessary to introduce a Higgs multiplet transforming as $(1,10^{\ast},0)$ \footnote{A redefinition of the fields in this multiplet has been introduced in \cite{PhysRevD.94.015007} (see Eq 75), this implies a scale factor in the mass terms of the gauge bosons coming from H when compared with our work. Some other important features of this version of the model are discussed in this reference}.
\be
\begin{array}{ccccc}
H=
\left(\begin{array}{ccccc}
H_{1}^0 & H_1^+ & H_2^0 & H_2^-\\
H_1^{+} & H_1^{++} & H_3^{+} & H_3^{0}\\
H_2^0 & H_3^+ & H_4^0 & H_4^-\\
H_2^{-}& H_3^0 & H_4^- & H_2^{--} 
\end{array}\right)
\sim(1,10^*,0),
\end{array}
\ee

with just three of their neutral fields developing a vaccum expectation value $\left<H_{2,3,4}^0\right>=v^{\p\p}$. It is important to mention that to preclude mixing among SM and the exotic quarks an extra multiplet $\eta^\prime$ must be introduced, transforming as $\eta$, but with different vacuum expectation value (VEV), $\langle\eta'\rangle=(0,0,v'/\sqrt{2},0)$. In this way we
have that the symmetry breaking pattern occurs according to:
\begin{align*}
  SU(4)_L \otimes U(1)_X\xrightarrow{\quad <\chi> \quad}SU(3)_L \otimes U(1)_{Z} \\
  \xrightarrow{\quad <\eta^\prime>, <\eta>, <\rho>, <H> \quad}
U(1)_{Q}.
\end{align*}

As for the gauge sector it is important to remember that the gauge group we are working is the $SU(4)_L \times U(1)_{X}$, it implies that there are $15$ $W^i_\mu$ ($i=1,...,15$) gauge bosons belonging to the $SU(4)_L$ group, and there is a singlet boson $B_\mu$ owned by the $U(1)_X$ group. The electric charge and interactions of the gauge bosons beyond the SM ones is determined by the chose we did for the electric charge operator \eqref{q}. In the diagonalization procedure we defined the physical charged gauge bosons as:  $-\sqrt{2}W^+=W^1-iW^2$, $-\sqrt{2}V^-_1=W^6-iW^7$, $-\sqrt{2}V_2^-=W^9-iW^{10}$, $-\sqrt{2}V_3^-=W^{13}-iW^{14}$, $-\sqrt{2}U^{--}=W^{11}-iW^{12}$ and $\sqrt{2}X^0=W^4+iW^5$. Notice the presence of three new single charged vector bosons $V^{\pm}_{1,2,3}$ and the existence of a doubly charged vector boson $U^{--}$. As we will show, these new vector bosons generates contributions to the anomalous magnetic moment. In the approximation that we worked, $V^{\pm}_{1,2}$ are degenerates and its contribution to the anomaly is the same, say $\Delta a_{\mu V1,V2}$. The $V^{\pm}_{3}$ will be heavier than the other two $V^{\pm}_{1,2}$ bosons, $M^{2}_{V3} \approx 2 \times M^{2}_{V1,V2}$, generating a $\Delta a_{\mu V3} \approx \frac{1}{2} \Delta a_{\mu V1,V2}$. The charged current interactions among the charged gauge bosons and the muon, relevant for the study of the anomaly, can be written as:

\begin{widetext}
\begin{align}
\label{chargedcurrent}
 \lag^{CC}\supset-\frac{g}{2\sqrt{2}} [\overline{\nu} \gamma^\mu (1-\gamma_5) \mu W^+_\mu + \overline{\nu^c} \gamma^\mu (1-\gamma_5) \mu V^+_{1_\mu}+\overline{\mu^c} \gamma^\mu (1-\gamma_5) \nu V^+_{2_\mu} + \overline{\mu^c} \gamma^\mu (1-\gamma_5) \nu^c V^+_{3_\mu}+\overline{\mu^c} \gamma^\mu (1-\gamma_5) \mu U^{++}_\mu] +H.C.,    
\end{align}
\end{widetext}
being $g$ the coupling constant of the electroweak group. 
As for the neutral sector, there are four neutral gauge bosons, the massless photon and three massive ones $Z_n$, with $Z_n = Z_N, Z, Z^\prime$ for $n=0,1,2$ respectively. To obtain the masses and the physical states in the neutral sector it is necessary to diagonalize the mass matrix in the basis $W^3,W^8,W^{15},B$,  given by

\begin{widetext}
\begin{equation}
\frac{g^2}{4}\left(
\begin{array}{llll}
v_1^2\!+\!u^2\!+2\!v''^2\, &\, \frac{1}{\sqrt3}(v_1^2\!-\!u^2\!-2\!v''^2)\, &\,
\frac{1}{\sqrt6}(v_1^2\!-\!u^2\!+\!4v''^2)\, &\, -2tu^2 \\
 \frac{1}{\sqrt3}(v_1^2\!-\!u^2\!-2\!v''^2)\, &\,
\frac{1}{3}(v_1^2\!+\!4v'^2\!+\!u^2\!+2\!v''^2)\, &
\frac{1}{3\sqrt2}(v_1^2\!-\!2v'^2\!+\!u^2\!-4\!v''^2)\, &\,
\frac{2}{\sqrt3}tu^2 \\
\frac{1}{\sqrt6}(v_1^2\!-\!u^2\!+\!4v''^2)\, &\,
\frac{1}{3\sqrt2}(v_1^2\!-\!2v'^2\!+\!u^2\!-\!4v''^2)
&\, \frac{1}{6}(v_1^2+v'^2+u^2+9V_\chi^2+8v''^2)\, &\,
\frac{2}{\sqrt6}t(u^2+3V_\chi^2) \\
-2tu^2\, &\, \frac{2}{\sqrt3}tu^2\, &\, \frac{2}{\sqrt6}t(u^2\!+\!3V_\chi^2)\,
&4t^2(u^2\!+\!V_\chi^2)
\end{array}
\right)
\label{zmass}
\end{equation}
\end{widetext}

where $t=\frac{\sin\theta_{W}}{\sqrt{1-4\sin^2\theta_{W}}}$, being $\theta_W$ the electroweak angle. In principle, the diagonalization of \eqref{zmass} has to be done numerically. However, an analytic solution can be found by setting $v_{1}=u=v''$ and $v'=V_{\chi}$, with $V_{\chi} \gg v_1$, yielding \cite{Pisano:1994tf}

\be M_n^2=\frac{g^2}{4}\lambda_nV_\chi^2 \label{mass} \ee
where $\lambda_{n}$ are constants given in the Appendix \ref{App1}. As for the charged gauge bosons, its masses are given by,

\begin{align} M_W^2=\frac{g^2}{4}(4v_1^2), \quad M_{V_1,2}^2=\frac{g^2}{4}(3v_1^2 + V_\chi^2),\\ \nonumber
M_{V_3}^2=\frac{g^2}{4}(2v_1^2 + 2V_\chi^2),\\ \nonumber
M_{U^{++}}^2=\frac{g^2}{4}(9v_1^2 + V_\chi^2).\end{align}

To calculate the contributions to the anomaly coming from the neutral vector sector, we must have in hand the neutral currents, which are given by: 

\be
\label{neutralcurrent}
{\cal L}^{NC} =-\frac{g}{2cos\theta_W}\left( \bar{\ell}_{L} \gamma^{\mu} \ell_L \alpha  + \bar{\ell}_{R} \gamma^{\mu} \ell_R \beta\right) Z_{n\mu},
\ee

where $\alpha$ and $\beta$ are couplings that are given explicitly in the appendix \ref{App2}. 

\begin{figure*}[ht] 
    \centering
    \includegraphics[scale=0.4]{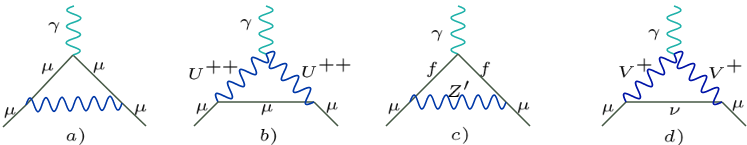}
    \caption{Feyman diagrams in order to contribute to the corrections to \am in the $SU(4)_L \otimes U(1)_X$ model, coming from the interactions with new gauge bosons: a) and b) doubly-charged gauge boson ($U^{++}$), c) neutral gauge boson ($Z^\prime$) and d) singly charged bosons ($V^+$).}
    \label{feynmandiagram1}
\end{figure*}

The corrections coming from charged and neutral scalars would be derived from the Yukawa Lagrangian:
\be
-\lag_{Yuk}=\supset\frac{1}{2} G_{ab} \overline{f^c_{aL}}f_{bL}H, 
\label{yukawa1}
\ee
where a,b= e, $\mu,\tau$.  These  scalars  interact  with  leptons through  the Yukawa Lagrangian in Eq \eqref{yukawa1}, meaning that they couple to leptons  proportionally  to  their  masses.  Hence,  their  contribution to $a_\mu$ will be suppressed.
Finally, \f \ref{feynmandiagram1} shows the Feynman diagrams of the interactions present in this model that contribute to the corrections \gmu.

\subsection{\textbf{\su} model without Exotic Electric Charges}
\label{model2}

In the \su model without Exotic Electric Charges \cite{PhysRevD.85.113010,Ponce_2004}, neutral heavy leptons are placed into the same multiplet that the left-handed charged leptons and neutrinos. All the right-handed fermions are singlets of $SU(4)_{L}$. In order to cancel all the quirial anomalies two left handed quark families must transform as 4-plets, and the other one as an anti-4-plet. So

\begin{align}
\begin{array}{ccccc}
f_{\alpha L}=
\left(\begin{array}{c}
\ell_{\alpha}\\
\nu_{\alpha}\\
N_{\alpha}\\
N_{\alpha}^{\prime}\end{array}\right)_{L}\sim(1,4^*,-1/2), \\  \\
Q_{iL}=\left(\begin{array}{c}
u_{i}\\
d_{i}\\
D_{i}\\
D_{i}^{\p}\end{array}\right)_{L} \sim(3,4,-1/6), \\  \\
Q_{3L}=\left(\begin{array}{c}
d_{3}\\
u_{3}\\
U\\
U^{\p}\end{array}\right)_{L}\sim(3,4^*,5/6),
\end{array}
\label{fermions}
\end{align}

where $\alpha$ is the flavor index $\alpha=1,2,3$, and $i=1,2$.

As for the right handed fields, they transform as:
\begin{equation}
\left(e_{\alpha R}\right)\sim(\mathbf{1,1},-2),    
\end{equation}

\begin{equation}
\begin{array}{c}
(d_{3R},(d_{iR}),(D_{iR}),(D_{iR}^{\prime})\sim(\mathbf{3},\mathbf{1},-2/3)\end{array}\label{Eq.2}\end{equation}

\begin{equation}
(u_{3R}),(u_{iR}),(U_{R}),(U_{R}^{\prime})\sim(\mathbf{3},\mathbf{1},4/3)\label{Eq.21}\end{equation}

The neutral heavy lepton masses are of the order $M_{N,N_{\alpha}^{\p}}\approx V_\chi /2$.

In the scalar sector, this model contains four scalar multiplets that develop a vaccum expectection value as follows \cite{PhysRevD.85.113010}
\begin{eqnarray*} 
\phi^T_1&=&(\zeta^0,\zeta^-_1,\zeta^-_2,\zeta^-_3) \sim[1,4,-3/2],\quad \left<\phi^T_1\right>=(v',0,0,0) \\ \nonumber
\phi^T_2&=&(\rho^+,\rho^0_1,\rho^0_2,\rho^0_3)  \sim[1,4,1/2],\quad \quad \left<\phi^T_2\right>=(0,v,0,0) \\ \nonumber
\phi^T_3&=&(\eta^+,\eta^0_1,\eta^0_2,\eta^0_3) \sim[1,4,1/2],\quad \quad \left<\phi^T_3\right>=(0,0,V,0)  \\ \nonumber
\phi^T_4&=&(\chi^+,\chi^0_1,\chi^0_2,\chi^0_3)\sim[1,4,1/2]. \quad \quad\left<\phi^T_4\right>=(0,0,0,V')\\ \label{scalars}
\end{eqnarray*}

This symmetry breaking pattern give masses to the fermions and gauge bosons of the model. The symmetry breaking occurs according to,
\begin{align*}
SU(4)_L\otimes  U(1)_X \xrightarrow{<\phi^T_4>} SU(3)_L\otimes U(1)_Z  \\
\xrightarrow{<\phi^T_3>} SU(2)_L\otimes U(1)_Y \xrightarrow{<\phi^T_1>,<\phi^T_2>}  U(1)_Q.
\end{align*}
the $SU(4)_L \otimes U(1)_X$ gauge group breaks down to $SU(3)_L \otimes U(1)_{Z}$ (3-3-1  model), by means of $<\phi^T_4>$ scalar boson. This latter group breaks down to $SU(2)_L \otimes U(1)_{Y}$ (gauge group of the SM) induced by $<\phi^T_3>$ scalar, and finally the last group breaks down to $U(1)_{Q}$, through two scalar bosons $<\phi^T_1>$ and $<\phi^T_2>$. 
In this work, we will work with the following simplifications for the VEVs  $V\sim V^\prime >> v\sim v^{\prime}$. Since $M^{2}_{W^{\pm}}=\dfrac{g^2}{2}(v^2+v^{\prime 2})$ we have that $\sqrt{v^2+v^{\prime2}} \approx 174Gev$ and then $v^{\prime} \sim 123 Gev$

For simplicity we will explicitly show only the interactions that contribute to the anomaly in this version of the 3-4-1 model, for a detailed explanation of the gauge sector in this model check please \cite{Ponce_2004}  

\begin{align}
\lag^{CC}_l \supset -\frac{g}{\sqrt{2}}\left[\bar{N}^{0}_{L}\gamma^\mu \mu_{L} K_{\mu}^{+} +\bar{N}_{L}^{0 \p}\gamma^\mu \mu_{L} X_{\mu}^+\right] + h.c., \nonumber \\
\lag^{NC} \supset \bar{\mu} \gamma^{\mu}[g_V-g_A\gamma^5]\mu Z^{\p}_{\mu}, 
\end{align}

being:

\begin{equation}
g_{V}=-\frac{g}{2\cos\theta_W}\frac{1-3\sin^{2}\theta_W}{\sqrt{3\cos^{2}\theta_W-1}},\nonumber    
\end{equation}

\begin{equation}
g_{A}=-\frac{g}{2\cos\theta_W}\frac{\cos^{2}\theta_W}{\sqrt{3\cos^{2}\theta_W-1}}    
\end{equation}

The mass eigenvalues of the gauge bosons we are interested here are:

\begin{align}
M_{K^{\pm}}^{2}=\dfrac{g^2}{2}(V^2+v^{\prime 2}),  M_{X^{\pm}}^{2}=\dfrac{g^2}{2}(V^{\p 2}+v^{\p 2}), \\ \nonumber \quad M_{Z^{\p}}^2=\dfrac{g^2}{4}V^2.
\label{massesN}
\end{align}

\subsection{\su model with Exotic leptons}
\label{model3}
In the \su model with Exotic leptons \cite{PhysRevD.77.035008}, instead of having neutral heavy leptons N, there are new exotic leptons denominated E. This lepton content is obtained by setting \textbf{$a=b=c=1$} in the electric charge operator \textbf{\eqref{Q}}. The left-handed fermion multiplets of the model are 

\begin{align}
\begin{array}{ccccc}
f_{\alpha L}=\left(\begin{array}{c}
\nu_{\alpha}\\
\ell_{\alpha}\\
E_{\alpha}^-\\
E_{\alpha}^{\p -}\end{array}\right)_{L}\sim(1,4,-3/4), \\ \\
Q_{iL}=\left(\begin{array}{c}
d_{i}^{\p}\\
u_{i}\\
U_{i}\\
U_{i}^{\p}\end{array}\right)_{L}\sim(3,4^*,5/12), \\ \\
Q_{3L}=\left(\begin{array}{c}
u_3\\
d_3\\
D_{3}\\
D_{3}^{\p}\end{array}\right)_{L}\sim(3,4,-1/12),
\end{array}
\end{align}
and the right-handed particles are singlets of the $SU(4)_{L}$ symmetry.

\begin{equation}
e_{\alpha L}^{c}\sim(1,1,1),  E_{\alpha L}^{c}\sim(1,1,1), E_{\alpha L}^{'c}\sim(1,1,1)
\end{equation}

\begin{equation}
\begin{array}{c}
(d^{c}_{3L}),(d^{c}_{iL}),(D^{c}_{iL}),(D^{\prime c}_{iL})\sim(3^*,1,+1/3),\\
(u^{c}_{3L}),(u^{c}_{iL}),(U^{c}_{L}),(U^{\prime c}_{L})\sim(3^*,1,-2/3)
\end{array}\label{Eq.20}\end{equation} where $i=1,2$ and $\alpha = 1,2,3$. To generates masses for the fermions and the gauge bosons, it is necessary the following scalar content:

\begin{align}
\phi^T_1=(\phi^0_1,\phi^+_1,\phi^{\p +}_1,\phi^{\p\p +}_1)\sim[1,4^*,3/4],\left<\phi^T_1\right>=(v_3,0,0,0),  \nonumber \\ 
\phi^T_2=(\phi^-_2,\phi^0_2,\phi^{\p 0}_2,\phi^{\p\p 0}_2)\sim[1,4^*,-1/4], \left<\phi^T_2\right>=(0,v',0,0),  \nonumber \\  
\phi^T_3=(\phi^-_3,\phi^0_3,\phi^{\p 0}_3,\phi^{\p\p 0}_3)\sim[1,4^*,-1/4],\left<\phi^T_3\right>= (0,0,V,0),  \nonumber \\ 
\phi^T_4=(\phi^-_4,\phi^0_4,\phi^{\p 0}_4,\phi^{\p\p 0}_4)\sim[1,4^*,-1/4], \left<\phi^T_4\right>=(0,0,0,V_\chi),
\label{vev3}
\end{align}

The symmetry breaking occurs in the same way as in the previous model. We assume $V \approx V_\chi>>v_3 \approx v^\prime$. 
The charged and neutral currents relevant for the anomaly are:

\begin{align}
\lag^{CC}\supset-\frac{g}{2\sqrt{2}} \left( \overline{\mu} \gamma^{\mu}(1-\gamma^5) E K^0_{\mu} + \overline{\mu} \gamma^{\mu} (1-\gamma^5) E^{\prime} X^0_{\mu}\right), \nonumber \\
\lag^{NC}\supset \bar{\ell} \gamma^{\mu} \left( g_V^\prime - g_A^\prime \gamma^5\right)\ell Z^{\prime},
\end{align}

where $K^0_{\mu}$, $X^0_{\mu}$ and $ Z^{\prime}$ are the only ones beyond SM gauge bosons contributing to \gmu, and

\be
g_V^\prime = \frac{g}{2\cos\theta_W}\frac{1/2 +\sin^2\theta_W}{\sqrt{2-3\sin^2\theta_W}};\nonumber 
\ee
\be
g_A^\prime= \frac{g}{2\cos\theta_W}\frac{\cos2\theta_W}{2\sqrt{2-3\sin^2\theta_W}}
\ee

After the neutral fields acquire its vaccum expectation value, as decribed in \eqref{vev3}, are generated the following mass terms for the bosons:

\begin{eqnarray}
M^2_{W^\pm}&=&\frac{g^2}{2}(v_3^2+v^{\p 2}), \nonumber \\
 M^2_{K^0}=\frac{g^2}{2}(v_3^2+V^2), \\ \nonumber
M^2_{X^0}&=&\frac{g^2}{2}(v_3^2+V_{\chi}^2),
\end{eqnarray}

being $g$ the coupling constant of the $SU(2)_L$ gauge group. As for the neutral gauge bosons, the $4\times4$ mass matrix has a zero eigenvalue corresponding to the photon. For the remainder $3\times3$ matrix we obtain the mass eigenvectors $Z_{\mu}$, $Z_{\mu}^{\prime}$ and $Z_{\mu}^{\prime \prime}$. In the approximation $V \sim V_{\chi}$, $Z_{\mu}^{\prime \prime}$ decouple from the other two, and it does not contributes to the anomaly, so it will be hereafter ignored. $Z_{\mu}$, $Z_{\mu}^{\prime}$
are still mixed, 
\begin{equation}
\frac{g_4^2}{C^2_W}\left(\begin{array}{cc} v_3^2 & \sqrt{2}\delta v^2_3 S_W \\
\sqrt{2}\delta v_3^2 S_W & \frac{2 \delta^2}{S^2_W}[v_3^2(S^4_W+C^4_W)+V^2 C^4_W]\end{array}\right),
\label{massmatrix}
\end{equation}

Here $g_4 = g$, $\delta=g_X/(2g)$, and $g_X$ is the gauge coupling constants of the $U(1)_X$ group. 

By diagonalizing this mass matrix we get the two physical neutral gauge bosons
\begin{eqnarray}\nonumber
Z_1^\mu&=&Z^\mu \cos\theta+Z^{\prime\mu} \sin\theta \; ,\\ \label{mixing}
Z_2^\mu&=&-Z^\mu \sin\theta+Z^{\prime\mu} \cos\theta, 
\end{eqnarray} 
where the mixing angle is given by
\begin{equation} \label{tan} \tan(2\theta) = \frac{2 \sqrt{2} \delta v_3^2 S^3_W}
{2 \delta^2[v_3^2(S^4_W+C^4_W)+V^2 C^4_W]-v_3^2 S^2_W}. 
\end{equation}

\section{RESULTS}
\label{results}

After presenting the key theoretical aspects of these three versions of the 3-4-1 model, now we will show our results. For each model, we calculated the individual contributions to the muon anomalous magnetic moment as function of the scale of symmetry breaking of the $SU(4)_{L}$ symmetry, $V_\chi$, and also, we computed the total contribution as function of $V_\chi$ to assess whether the model accommodates the anomaly or not. The analytical expressions used in this work are shown in the appendix \ref{App3}, and were taken from \cite{Lindner:2016bgg}. Besides, we provide the numerical codes we used to derive our results \cite{Mathematicacode}. As previously mentioned the corrections to \amu coming from scalar fields are suppressed by their couplings, which are proportional the muon mass, for this reason, we will ignore them in our calculations. In section \ref{AP}, we discuss how one can make the scalar corrections sizeable and meaningful to the g-2 anomaly. We will draw our conclusions having in mind lower mass bounds stemming from collider searches for new gauge bosons.

\begin{figure*}[!htp]
    \centering
    \includegraphics[scale=0.7]{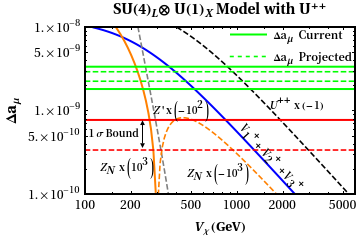}
    \includegraphics[scale=0.7]{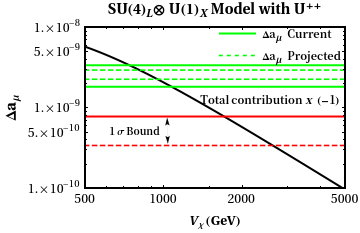}
    \caption{Individual contributions (left-panel) and total contribution (right-panel) to \amu as a function of the scale of symmetry breaking $V_\chi$, coming from the \su model with doubly charged gauge boson. The current and projected experimental bounds are exhibited with thick and dashed green lines respectively. The 1 $\sigma$ current and projected  error are displayed as thick and dashed red lines respectively}
    \label{1model}
\end{figure*}

\begin{figure*}[!htp]
    \centering
    \includegraphics[scale=0.7]{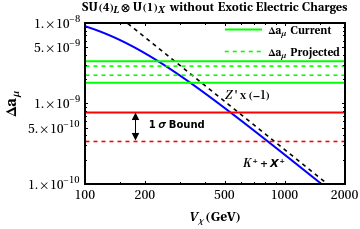}
    \includegraphics[scale=0.7]{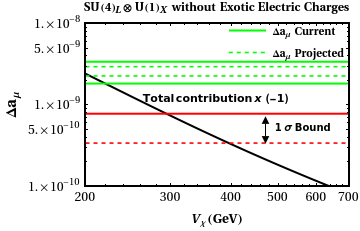}
    \caption{ Individual contributions (left-panel) and total contribution (right-panel) to \amu as a function of the scale of symmetry breaking $V_\chi$, coming from the \su model without exotic electric charges for $M_N$=$M_{N^\prime}=10$GeV. The current and projected experimental bounds are exhibited with thick and dashed green lines respectively. The 1 $\sigma$ current and projected  error are displayed as thick and dashed red lines respectively.}
    \label{2model}
\end{figure*}

\begin{figure*}[!htp]
    \centering
    \includegraphics[scale=0.7]{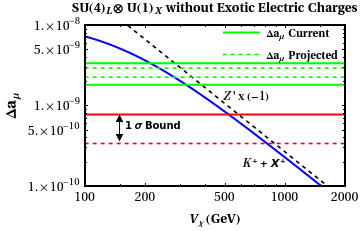}
    \includegraphics[scale=0.7]{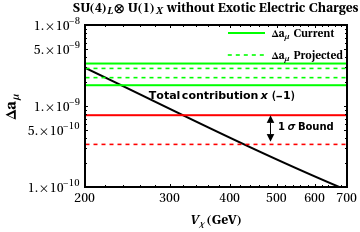}
    \caption{Individual contributions (left-panel) and total contribution (right-panel) to \amu as a function of the scale of symmetry breaking $V_\chi$, coming from the \su model without exotic electric charges for $M_N$=$M_{N^\prime}=100$GeV. The current and projected experimental bounds are exhibited with thick and dashed green lines respectively. The 1 $\sigma$ current and projected  error are displayed as thick and dashed red lines respectively.}
    \label{3model}
\end{figure*}

\begin{figure*}[!htp]
    \centering
    \includegraphics[scale=0.7]{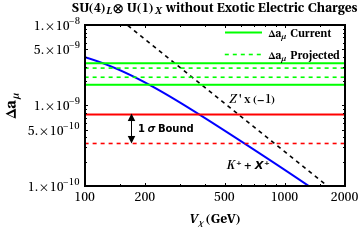}
    \includegraphics[scale=0.7]{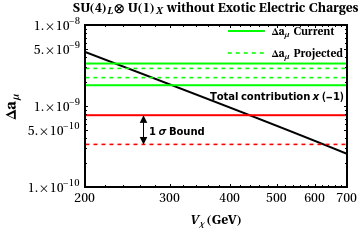}
    \caption{Individual contributions (left-panel) and total contribution (right-panel) to \amu as a function of the scale of symmetry breaking $V_\chi$, coming from the \su model without exotic electric charges for $M_N$=$M_{N^\prime}=1 TeV$. The current and projected experimental bounds are exhibited with thick and dashed green lines respectively. The 1 $\sigma$ current and projected  error are displayed as thick and dashed red lines respectively.}
    \label{31model}
\end{figure*}

\begin{figure*}[!htp]
    \centering
    \includegraphics[scale=0.7]{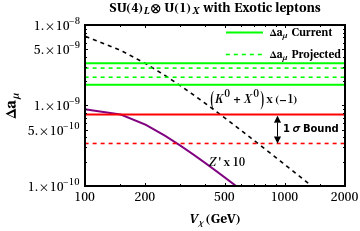}
    \includegraphics[scale=0.7]{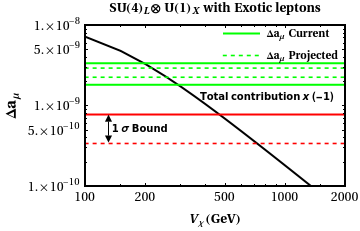}
    \caption{Individual contributions (left-panel) and total contribution (right-panel) to \amu as a function of the scale of symmetry breaking $V_\chi$, coming from the \su model with exotic leptons for $M_{E_1}$=$M_{E_2}=10$GeV. The current and projected experimental bounds are exhibited with thick and dashed green lines respectively. The 1 $\sigma$ current and projected  error are displayed as thick and dashed red lines respectively.}
    \label{4model}
\end{figure*}

\begin{figure*}[!htp]
    \centering
    \includegraphics[scale=0.7]{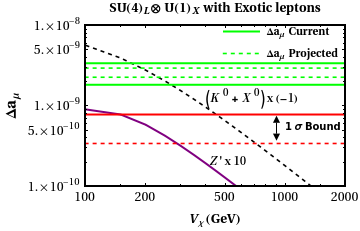}
    \includegraphics[scale=0.7]{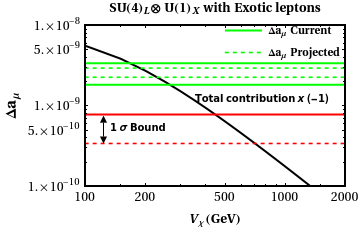}
    \caption{Individual contributions (left-panel) and total contribution (right-panel) to \amu as a function of the scale of symmetry breaking $V_\chi$, coming from the \su model with exotic leptons for $M_{E_1}$=$M_{E_2}=100$GeV. The current and projected experimental bounds are exhibited with thick and dashed green lines respectively. The 1 $\sigma$ current and projected  error are displayed as thick and dashed red lines respectively.}
    \label{6model}
\end{figure*}

\begin{figure*}[!htp]
    \centering
    \includegraphics[scale=0.7]{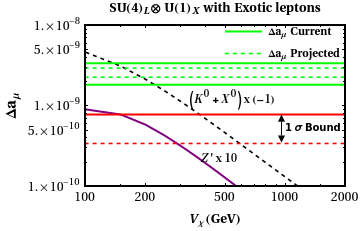}
    \includegraphics[scale=0.7]{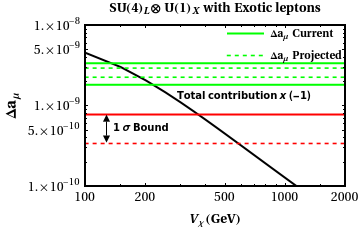}
    \caption{Individual contributions (left-panel) and total contribution (right-panel) to \amu as a function of the scale of symmetry breaking $V_\chi$, coming from the \su model with exotic leptons for $M_{E_1}$=$M_{E_2}=1 TeV$. The current and projected experimental bounds are exhibited with thick and dashed green lines respectively. The 1 $\sigma$ current and projected  error are displayed as thick and dashed red lines respectively.}
    \label{5model}
\end{figure*}

\begin{table*}[htp]
    \centering
    \begin{tabular}{|c|c|c|c|c|c|c|} \hline
   Bounds&$V_\chi$(GeV)&$M_{V_1^+,V_2^+}$(GeV)&$M_{V_3^+}$(GeV)&$M_{U^{++}}$(GeV)&$M_{Z^{\p}}$(GeV)&$M_{Z_{N}}$(GeV)\\ \hline \hline
   $\sigma$ current & $\geq1700$&$\geq557$ &$\geq783$ & $\geq565$ &$\geq675$&$\geq2048$\\ \hline
   $\sigma$ projected&$\geq2615 $&$\geq853$& $\geq1203$& $\geq858$& $\geq1038$&$\geq3150$\\
    \hline  \hline
    \end{tabular}
    	\caption{Lower bounds on the scale of symmetry breaking $V_{\chi}$ and masses of the gauge bosons for the \su model with doubly charged gauge boson, for details see the text.}
	\label{Table00}
\end{table*}

\begin{table*}[ht]
    \centering
    \begin{tabular}{|c||c|c|c|c|} \hline
    Models& Bounds&$M_i=10GeV$&$M_i=100GeV$&$M_i=1000GeV$ \\ \hline \hline
      	\multirow{1}{6.5 cm}{\su without exotic electric charges} 
      	&\multirow{2}{2 cm}{$\sigma$ current}
      	&$V_X > 293GeV$ &$V_X>318 GeV$ & $V_X>438GeV$ \\ \cline{3-5}
      	&& $M_{K^+, X^+}>146GeV$  &$M_{K^+, X^+}>157GeV$ &$M_{K^+, X^+}>209GeV$\\ \cline{3-5}
      	&& $M_{Z^\p}>95GeV$  &$M_{Z^\p}>103GeV$ &$M_{Z^\p}>142GeV$\\ 
	\cline{2-5} 
	&\multirow{2}{2 cm}{$\sigma$ projected}
&$V_X>394GeV$ &$V_X>426GeV$&$V_X>624GeV$\\ \cline{3-5}
&& $M_{K^+, X^+}>190GeV$  &$M_{K^+, X^+}>204GeV$ &$M_{K^+, X^+}>292GeV$\\ \cline{3-5}
      	&& $M_{Z^\p}>128GeV$  &$M_{Z^\p}>138GeV$ &$M_{Z^\p}> 203GeV$\\ 
	\cline{1-5} 
		\multirow{1}{6.5 cm}{\su with Exotic Leptons}  
      	&\multirow{2}{2 cm}{$\sigma$ current}
      &$V_X >466GeV$ &$V_X>445GeV$&$V_X>365GeV$\\ \cline{3-5}
      	&& $M_{K^0 , X^0}>222GeV$   &$M_{K^0 , X^0}>212GeV$  &$M_{K^0 , X^0}>177GeV$ \\ 
	\cline{3-5} 
		&& $M_{Z^\p}>340GeV$  &$M_{Z^\p}>327GeV$ &$M_{Z^\p}>273GeV$\\ 
	\cline{2-5}
	&\multirow{2}{2 cm}{$\sigma$ projected}
	&$V_X>719 GeV$ &$V_X>702GeV$&$V_X>578GeV$ \\ \cline{3-5}
	&& $M_{K^0 , X^0}>335GeV$   &$M_{K^0 , X^0}>328GeV$  &$M_{K^0 , X^0}>272GeV$ \\ 
	\cline{3-5} 
		&& $M_{Z^\p}>515GeV$  &$M_{Z^\p}>503GeV$ &$M_{Z^\p}>418GeV$\\ 
	\cline{1-5}
	\hline
      \hline
    \end{tabular}
    	\caption{Lower bounds on the scale of symmetry breaking $V_{\chi}$ and masses of the gauge bosons for the \su model without exotic electric charges, for three different mass values of the neutral heavy leptons $N, N^{\prime}$, (Up panel). Lower bounds on the scale of symmetry breaking $V_{\chi}$ and masses of the gauge bosons for the \su model with exotic electric charges, for three different mass values of the exotic leptons $E_{1}, E_{2}$, (down panel).}
	\label{Table0}
\end{table*}

\subsection{\su model with doubly charged gauge boson}

First of all, let us remind the readers that in this version of the 3-4-1 model, we are working with the simplifications $v_{1}=u=v''$ and $v'=V_{\chi}$, with $V_{\chi} \gg v_1$, and that $v_{1}\sim123Gev$. In \f \ref{1model} we show the individual contributions to $\Delta a_{\mu}$ as a function of the scale of symmetry breaking of the $SU(4)_{L}$ group, $V_\chi$. To assess whether this model accommodates the muon anomalous magnetic moment, we also show the total contribution of the model as function of $V_\chi$. We verify that the new neutral gauge bosons $Z^\prime$ and $Z_N$ have small and negative contributions to \amu. This occurs because in the limit $M_{Z^{\prime},Z_{N}} \gg m_{\mu}$,  $m_\mu$ being the muon mass, their contributions to \amu are proportional to $g^2_V-5g^2_A$. The  singly charged gauge bosons ($V^{+}_ 1 + V^{+}_ 2 + V^{+}_ 3$) corrections are positive, but not enough to compensate for the larger and negative contribution of the doubly charged gauge boson $U^{++}$. The sign of the contribution of the doubly charged gauge boson is due to the nature of its coupling with the muons. As was proved in \cite{Lindner:2016bgg}, the $U^{++}$ couples to muons axially, its vector coupling is null. As the total contribution to the anomaly is negative, this model can not explain it, therefore, we can simply demand that the total contribution be smaller than the error bar. From the $1\sigma$ current bound, we obtain the lower limit $V_\chi> 1700$~GeV; and $V_\chi> 2615$~GeV from $1\sigma$ projected bound (see \f \ref{1model}). In accordance with the equation \eqref{mass}, these bounds translate into $M_{Z'} \ge 675$~GeV and $M_{Z'} \ge 1038$~GeV. These bounds are weaker than the LHC one for the $Z^{\prime}$ mass, which lies around $3.7$~TeV \cite{Nepomuceno:2019eaz} if the $Z^\prime$ boson decays exclusively into charged leptons. When exotic decays are present this limit weakens, but is yet stronger than the g-2 ones. It is important to emphasize that although this LHC limit has been derived for the minimal 3-3-1 model, it apply to our model also. This is because the 3-3-1 models are the low energy realization of 3-4-1 models. Each 3-3-1 version inherits the physical properties of some of the 3-4-1 models. Hence, the 3-4-1 we are working on it has the minimal 3-3-1 model as its low energy realization. As for the doubly charged gauge boson, our g-2 study translates into the following bounds, $M_{U^{++}} > 565$~GeV and $M_{U^{++}} > 858$~GeV using the $1\sigma$ current and projected bound, respectively. The most updated bound on the mass of this doubly charged gauge boson is $M_{U} \ge 1.2$~TeV \cite{Nepomuceno:2019eaz}. For the singly charged bosons $V_1$ and $V_2$, using the $1\sigma$ current bound we obtain $M_{V1,V2} > 556$~GeV and using the  $1\sigma$ projected bound we obtain $M_{V1,V2} > 852$~GeV. As for the singly charged boson $V_3$, using the $1\sigma$ current bound  we obtain $M_{V3} > 783$~GeV and using the  $1\sigma$ projected bound we obtain $M_{V3} > 1203$~GeV. The most updated bound on the mass of these singly charged gauge bosons reads $M_{V} \ge 850$~GeV \cite{Nepomuceno:2019eaz}. Thus, despite not being able to address g-2, our study led to the strongest lower mass bound on the singly charged gauge boson.


\subsection{\su without Exotic Electric Charges}

In this model without Exotic Electric Charges we calculated the contributions to the anomaly by working with the following simplifications for the VEVs  $V\sim V^\prime >> v\sim v^{\prime}$ and $v^{\prime} \sim 123$~GeV.
In \fs \ref{2model}, \ref{3model} and \ref{31model} we show the individual contributions to $\Delta a_{\mu}$ as a function of the scale of symmetry breaking, $V_\chi$, and as before, to assess whether this model accommodates the anomaly, we also show the total contribution of the model as function of $V_\chi$, for three different mass values of the neutral heavy leptons, $M_{N},M_{N^{\prime}}=10 GeV$, $M_{N},M_{N^{\prime}}=100 GeV$ and $M_{N},M_{N^{\prime}}=1 TeV$ respectively. As we can see, for the three mass values of the neutral heavy leptons, the contribution of the neutral $Z^{\prime}$ is negative and greater than the positive $K^{+}+X^{+}$ contribution, producing a negative total contribution in all cases. Due to these negative contributions, we conclude that this model cannot explain $\Delta a_{\mu}$. As before, we just enforce that the total contribution be smaller than the error. By using the $1\sigma$ current and projected bounds, we derived the lower limits $V_\chi> 293$~GeV, and $V_\chi> 394$~GeV respectively, in the case $M_{N},M_{N^{\prime}}=10$~GeV. For the case $M_{N},M_{N^{\prime}}=100$~GeV, we derived the lower limit $V_\chi> 318$~GeV by using the $1\sigma$ current bound, and $V_\chi> 426$~GeV from $1\sigma$ projected bound. Finally, for the case $M_{N},M_{N^{\prime}}=1$~TeV, we get $V_\chi> 438$~GeV by using the $1\sigma$ current bound, and $V_\chi> 624$~GeV from $1\sigma$ projected bound. The lower limits on the masses of the $K^{+}, X^{+}$ and $Z^{\prime}$ bosons for the different $M_{N},M_{N^{\prime}}$ values are shown in table II. As in the previous case, where the minimal 3-3-1 model inherits the physical properties of our 3-4-1 model with doubly charged gauge boson, in this case, the 3-3-1 model that inherits the properties of our 3-4-1 model without exotic electric charges, is the 3-3-1 LHN  \cite{PhysRevD.83.065024, PhysRevD.86.073015}. The collider bounds derived for this 3-3-1 version are similar to the bounds derived for the 3-3-1 RHN model \cite{PhysRevD.54.4691, PhysRevD.53.437}. For the 3-3-1 RHN model the collider bound has been derived as $M_{Z'} \ge 4$~TeV \cite{Lindner:2016bgg}, that would translate into a lower bound $V_{\chi} \ge 12$~TeV. In the 3-3-1 LHN model this bound can be weakened if we consider additional decay modes, as exotic quarks and the neutral heavy leptons itself. Including these decay modes, the bound is read now as $M_{Z'} \ge 2$~TeV or $V_{\chi} \ge 6$~TeV \cite{deJesus:2020ngn}, still strongest than the bounds derived from the g-2.

\subsection{\su with Exotic leptons}

For the 3-4-1 model with exotic leptons the symmetry breaking pattern is such that $V \approx V_\chi>>v_3 \approx v^\prime =123$~GeV. 
In \fs \ref{4model}, \ref{6model} and \ref{5model} we show the total and individual contributions to $\Delta a_{\mu}$ as a function of the scale of symmetry breaking, $V_\chi$, for three different mass values of the exotic heavy leptons, $M_{E_1},M_{E_2}=10 GeV$, $M_{E_1},M_{E_2}=100$~GeV and $M_{E_1},M_{E_2}=1$~TeV respectively. As we can see, for all mass values of the exotic leptons, the contribution of the bosons $K^{0}+X^{0}$ is negative and greater than the low and positive contribution of the $Z^{\prime}$ boson.  Therefore, the total contributions are always negative, and for this reason this version of the 3-4-1 model can not explain \amu. What we have left is to derive some constraints demanding that the total contribution be less than the error, as we did for the other models. By using the $1\sigma$ current bound, we derived the lower limit $V_\chi> 466$~GeV, and $V_\chi> 719$~GeV from $1\sigma$ projected bound, for the case $M_{E_1},M_{E_2}=10$~GeV. For the case $M_{E_1},M_{E_2}=100$~GeV, we derived the lower limit $V_\chi> 445$~GeV by using the $1\sigma$ current bound, and $V_\chi> 702$~GeV from $1\sigma$ projected bound. Finally, for the case $M_{E_1},M_{E_2}=1$~TeV, we derived the lower limit $V_\chi> 365$~GeV by using the $1\sigma$ current bound, and $V_\chi> 578$~GeV from $1\sigma$ projected bound.
The lower limits on the masses of the $K^{0}$, $X^{0}$ and $Z^{\prime}$ bosons for the different $M_{N},M_{N^{\prime}}$ values are shown in table II. One can clearly see that even for the case where $V_\chi> 719$~GeV (the greater value of $V_{\chi}$), the lower mass limit $M_{Z'} \ge 515$~GeV is weak and far from the existing collider bound.

\begin{figure*}[!htp]
    \centering
    \includegraphics[scale=0.9]{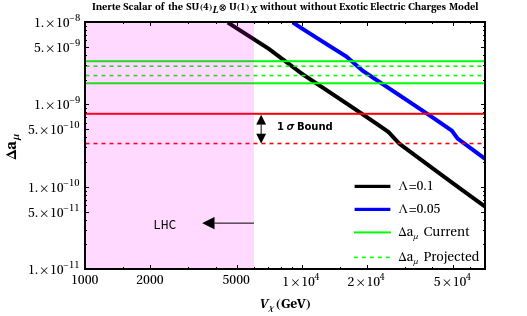}
    \caption{Inert scalar contribution to \amu as a function of the scale of symmetry breaking $V_\chi$, for the \su model without exotic electric charges. The current and projected experimental bounds are exhibited with thick and dashed green lines respectively. The 1 $\sigma$ current and projected  error are displayed as thick and dashed red lines respectively. For when $\Lambda=0.1$  the model accommodate the anomaly for $8Tev \leq V_{\chi} \leq 11 Tev$, and for when $\Lambda=0.05$  the model accommodate the anomaly for $16Tev \leq V_{\chi} \leq 22 Tev$.}
    \label{newmodel}
\end{figure*}

\subsection{Alternative paths}
\label{AP}

We have demonstrated that all three models based on 3-4-1 symmetry we investigated in this work cannot accommodate the muon anomalous magnetic moment. Therefore, a natural question arises. Can we make these models simultaneously consistent with the g-2 anomaly and the existing collider bounds? The answer is yes. We remind the reader that the scalar contributions to g-2 are dwindled because they couple to muons proportional to the muon mass (see for example Eq 11 from \cite{deJesus:2020ngn}). As was pointed out in \cite{deJesus:2020ngn} and most recently in \cite{jesus2020vectorlike}, a way to salvage these extended gauge sectors is by introducing inert scalars, or heavy vector-like charged leptons plus singlet scalars (in the case the vector-like heavy lepton be a singlet of the gauge symmetry). Here we will address how the first of these ideas can be implemented in the 3-4-1 model without exotic electric charges. As was shown in the section \ref{model2}, in this model we dispose of four scalars multiplets transforming as $\sim 4$ by the $SU(4)_{L}$ symmetry. All of them develop vacuum expectation value, and for this reason they are not inert scalars and their interactions with fermions are proportional to the fermion mass. We now introduce a new inert scalar $\phi_{1}^{\prime}$, which is a replica of $\phi_{1}$ \footnote{and to avoid the proliferation of scalar particles we could eliminate $\phi_{2}$, that has been introduced in \cite{PhysRevD.85.113010} to implement the See-Saw mechanism, and then look for a new way to generate neutrino masses in the model}. The Yukawa lagrangian that generates contributions to the muon anomaly is now:

\begin{equation}
\lag^{Y}=f_{ \alpha ,\beta  }\bar{f}_{\alpha L}  \phi _{ 1 }^{\star}\ell _{ \beta R }+\lambda _{ \alpha ,\beta  }\bar{f}_{\alpha L} \phi _{ 1 }^{ \prime  \star}\ell _{ \beta R }+H.C.
\label{yukawa}
\end{equation}

For the case of the $\phi_{1}$ scalar, the interactions are:
\[f_{\alpha, \beta}\left(\bar{\ell}_{\alpha L}\zeta^{0 \star}\ell_{\beta R}+\bar{\nu}_{\alpha L}\zeta_{1}^{+}\ell_{\beta R}+\bar{N}_{\alpha L}\zeta_{2}^{+}\ell_{\beta R}+\bar{N^{\prime}}_{\alpha L}\zeta_{3}^{+}\ell_{\beta R}\right).\] As the $\phi_{1}$ multiplet develops vacuum expectation value, its interactions with fermions are proportional to the fermion mass. For simplicity let us check the diagonal interaction $f_{2,2}\bar{\mu}_{L}R_{\zeta}\mu_{R}$. In this case $f_{2,2} \sim \frac{m_{\mu}}{v'} \sim 10^{-4}$, and $R_{\zeta}$ is the mass eigenvector obtained after the diagonalization procedure in the neutral scalar sector. The other three diagonal interactions among the muon, singly charged scalars, and neutral leptons, share the same $f_{2,2}$ coupling. As $f_{2,2} \sim 10^{-4}$ and $\Delta a_{\mu}$ is proportional to the square of this coupling constant, all the conributions coming from $\phi_{1}$ are irrelevants \footnote{For details about the individual contributions to the anomaly coming from $\phi_{1}$, see the appendix \ref{NS}}. As for the interactions $\lambda_{\alpha, \beta}\left(\bar{\ell}_{\alpha L}\zeta^{\prime 0 \star}\ell_{\beta R}+\bar{\nu}_{\alpha L}\zeta_{1}^{\prime +}\ell_{\beta R}+\bar{N}_{\alpha L}\zeta_{2}^{\prime +}\ell_{\beta R}+\bar{N^{\prime}}_{\alpha L}\zeta_{3}^{\prime +}\ell_{\beta R}\right)$ coming from the $\phi_{1}^{\prime}$ inert scalar, the situation is very different. As $\phi_{1}^{\prime}$ is inert, the diagonal coupling $\lambda_{2, 2}$ (as well as any other $\lambda_{\alpha, \beta}$) is not proportional to the fermion mass. Again let us check the first of the interactions, $\lambda_{2, 2}\bar{\mu}_{L}\zeta^{\prime 0 \star}\mu_{R}$, this coupling (and the others
proportional to $\lambda_{2, 2}$, among the muon, charged scalars and neutral leptons, like $\sim \lambda_{2, 2}\bar{N}_{2 L}\zeta_{2}^{\prime +}\mu_{R}$) contributes to the anomaly. The mass of the $\phi^{\prime}_{1}$ scalar comes from the potential terms $\sim \Lambda\phi_{3}^{\dagger}\phi_{3}\phi_{1}^{\prime \dagger}\phi_{1}^{\prime}+\Lambda\phi_{4}^{\dagger}\phi_{4}\phi_{1}^{\prime \dagger}\phi_{1}^{\prime}$. After $\phi_{3}, \phi_{4}$ develop vacuum expectation value, the scalar $\phi_{1}^{\prime}$ gains mass  proportional to $M_{\phi_{1}^{\prime}} \sim \Lambda V_{\chi} $. By setting the parameter $\lambda_{2,2}=1$ and using the equation \eqref{DeltaNeutralScalar} of appendix \ref{NS}, we calculated the main contribution of $\phi_{1}^{\prime}$ to the anomaly for two different values of $\Lambda$ (see figure \ref{newmodel}) \footnote{The main contribution comes from the $\zeta^{\prime 0 \star}$ particle, as explained in the appendix \ref{NS}}. For when $\Lambda=0.1$  the model accommodate the anomaly for $8Tev \leq V_{\chi} \leq 11$~ TeV, and for when $\Lambda=0.05$  the model accommodate the anomaly for $16 {\rm TeV} \leq V_{\chi} \leq 22$~TeV.

\section{conclusions}

We have revisited the muon anomalous magnetic moment in the context of the $SU(3)_C \times SU(4)_L \times  U(1)_X$ symmetry. We have used updated and correct analytical expressions to obtain the corrections to g-2. Our numerical results differ from previous works, but the overall conclusions are basically the same, that is, the models can not accommodate the g-2 anomaly and the bounds on the scale of symmetry breaking rising from g-2 are weaker than the ones coming from collider searches. A key point overlooked in the past is the contribution from the doubly charged gauge boson ,$U^{++}$, which is negative and not positive as previously assumed.  Moreover, we also derived new bounds on the scale of symmetry breaking of these models in the light of the upcoming results of the g-2 experiment at FERMILAB. We concluded that none of these models can accommodate the discrepancy among the theory and the current and projected experimental results. As a final contribution of this work, we presented a way via inert scalars to salvage these models. Basically, we look for inert scalars multiplets that couple to leptons multiplets in an invariant way. These inert multiplets are replicas of the scalar multiplets that couple with leptons and generates its masses. That allowed us to address g-2 while being consistent with current collider bounds.

\section{ACKNOWLEDGEMENTS}

DC  is partly supported by the Brazilian National Council for Scientific and Technological Development (CNPq), under grants 436692/2018-0. YSV acknowledges the financial support from CAPES under grants 88882.375870/2019-01. YMOT acknowledges the financial support from CAPES under grants 88887.485509/2020-00.

\begin{widetext}

\section{Appendix}
\label{appendix}
\subsection{Masses of the gauge bosons: \su model with doubly charged gauge boson}
\label{App1}
Coupling constants that appear in the neutral boson mass equation \e \ref{mass}.

\begin{mathletters}
\label{lambda}

\begin{equation}
\lambda_{n}=\frac{1}{3}\left[A+2\left(A^2+3B\right)^{\frac{1}{2}}
\cos\left(\frac{2n\pi+\Theta}{3} \right)
 \right],
\label{ln}
\end{equation}
where,
\begin{equation}
A=3+4t^{2}+(7+4t^{2})a^{2}, B=-2\left[1+3t^{2}+2(4+9t^{2})a^{2}\right],
\label{ab}
\end{equation}
\begin{equation}
C=8(1+4t^2)a^2,\quad \Theta=\arccos\left[
\frac{2A^3+9AB+27C}{2(A^2+3B)^{\frac{3}{2}}}\right],
\label{c}
\end{equation}
\end{mathletters}
with $a\equiv v_1/V_{\chi}$, and $t=g^{\prime}/g$.

\subsection{Vector and axial couplings: \su model with doubly charged gauge boson}
 \label{App2}
The derivation of the vector and axial couplings is a bit tedious, our results agree with Ref.\cite{Pisano:1994tf}. Defining,

\begin{equation}
Z_{n\mu}\approx x_nW^3_\mu+y_nW^8_\mu+z_nW^{15}_\mu+w_nB_\mu,
\label{autovec}
\end{equation}with
\begin{mathletters}
\label{xyzw}
\begin{equation}
x_n=-\frac{2a^2}{t}\frac{1-3t^{2}+(1-t^{2})a^{2}-(1-2t^{2})\lambda_n}{D_{n}(t,a)} w_{n},
\label{1a}
\end{equation}
\begin{eqnarray}
y_n&=&\frac{1}{\sqrt3 t}\frac{2(2+t^{2})a^{2}-10a^{4}t^{2}-\left[1+(1-4t^{2})a^{2}\right]\lambda_n}{D_{n}(t,a)}w_n,
\label{2a}
\end{eqnarray}
\begin{equation}
z_n=\frac{1}{\sqrt6t} \frac{8(2+t^{2})a^{2}+4(3+2t^{2})a^{4}-4\left[1+2(2+t^{2})a^{2}\right]\lambda_{n}+
3\lambda_{n}^{2}}{D_{n}(t,a)} w_{n},
\label{3a}
\end{equation}
\begin{equation}
w^2_n=\frac{1}{1+x_n^2/w_n^2+y_n^2/w_n^2+z_n^2/w_n^2},
\label{4a}
\end{equation}
\end{mathletters}
and finally
\begin{equation}
D_{n}(t,a)=2(7+5a^{2})-(3+13a^{2})\lambda_{n}+2\lambda_{n}^{2}.
\end{equation}

The vector and axial couplings can be derived from  the Lagrangian
 \begin{equation}
{\cal L} = -\frac{g}{2C_W} \left( \bar{l_L} \gamma^{\mu} l_L \alpha  + \bar{l_R} \gamma^{\mu} l_R \beta\right) Z_{n\mu},
\end{equation}with

 \begin{equation}
\alpha=-c_W\left(-x_n+\frac{1}{\sqrt3}
y_n+\frac{1}{\sqrt6}z_n+\frac{4}{3}w_nt\right) +\frac{4}{3}c_W w_nt,\quad \beta =-\frac{3}{\sqrt6}z_n.
\label{leptons}
\end{equation}with $Z_0=Z_N, Z_1=Z,Z_2=Z^{\prime}$.

\subsection{Analytical Expressions for the Muon Magnetic Moment} 
\label{App3}

The analytical expressions for all the contributions to \gmu that have been used in the present manuscript were taken from \cite{Lindner:2016bgg}. 

\subsubsection{Charged Fermion -- Doubly Charged Vector Boson}

The contribution to the muon anomalous magnetic moment coming from a doubly charged vector boson is derived from the following general lagrangian

\begin{equation}
  \mathcal{L}_\textrm{int} = g^{ ij }_{ V }U^{ ++ }_{ \mu  }\, \overline { \ell _{ i }^{ C } } \gamma ^{ \mu  }\, \ell _{ j }+g^{ ij }_{ A }U^{ ++ }_{ \mu  }\, \overline { \ell _{ i }^{ C } } \gamma ^{ \mu  }\gamma ^{ 5 }\ell _{ j }+ \mathrm{h.c.}
\end{equation}

In this case, \gmu is given by:

\begin{align}
  \Delta a_\mu\left(U^{++}\right) &= \frac { 8 }{ 8\pi ^{ 2 } } \frac { m_{ \mu  }^{ 2 } }{ m_{ U }^{ 2 } } \int _{ 0 }^{ 1 }{ d } x\frac { g_{ V }^{ 2 }P_{ 1 }^{ + }(x)+g_{ A }^{ 2 }P_{ 1 }^{ - }(x) }{ \epsilon _{ f }^{ 2 }\lambda ^{ 2 }(1-x)\left( 1-\epsilon _{ f }^{ -2 }x \right) +x }  \nonumber\\
  &+\frac { 4 }{ 8\pi ^{ 2 } } \frac { m_{ \mu  }^{ 2 } }{ m_{ U }^{ 2 } } \int _{ 0 }^{ 1 } { d }x\frac { g_{ V }^{ 2 }P_{ 2 }^{ + }(x)+g_{ A }^{ 2 }P_{ 2 }^{ - }(x) }{ (1-x)\left( 1-\lambda ^{ 2 }x \right) +\epsilon _{ f }^{ 2 }\lambda ^{ 2 }x } , \label{eq:Delta_a_Doubly_Vector} 
\end{align}

with $\epsilon_f \equiv \frac{m_f}{m_\mu}$ and $\lambda \equiv \frac{m_\mu}{m_U}$. This contribution arises in the first model, the model with doubly charged gauge boson, where ${ m }_{ f }={ m }_{ \mu  }$. The polynomials inside the integrals are defined as:

\begin{equation}
P_{ 1 }^{ \pm  }=-2x^{ 2 }(1+x\mp 2\epsilon _{ f })+\lambda ^{ 2 }x(1-x)(1\mp \epsilon _{ f })^{ 2 }(x\pm \epsilon _{ f }),
\label{p1poly}
\end{equation}

and 

\begin{eqnarray}
{ P }_{ 2 }^{ \pm  }=2x\left( 1-x \right) \left( x-2\pm 2{ \epsilon  }_{ f } \right) +{ \lambda  }^{ 2 }{ x }^{ 2 }{ \left( 1\mp { \epsilon  }_{ f } \right)  }^{ 2 }\left( 1-x\pm { \epsilon  }_{ f } \right) 
\label{p2poly}
\end{eqnarray}

As was pointed out in the text, the $U^{++}$ couples to muons axially, its vector coupling is null. Taking this into account, and the fact that $\epsilon_{f}=1$, the polynomials that enter in the contribution are reduced to:

\begin{equation}
{ P }_{ 1 }^{ - }\left( x \right) =-2{ x }^{ 2 }(x+3)+4{ \lambda  }^{ 2 }x(1-x){ (x-1) },
\label{p1minus}
\end{equation}

\begin{equation}
{ P }_{ 2 }^{ - }\left( x \right) =2{ x }(1-x)(x-4)-4{ \lambda  }^{ 2 }{ x }^{ 3 } 
\label{p2minus}
\end{equation}

From the integral \eqref{eq:Delta_a_Doubly_Vector} and using the polynomials \eqref{p1minus} and \eqref{p2minus} above, we have obtained the contribution of the $U^{++}$ boson. We remember the readers that these expressions \eqref{eq:Delta_a_Doubly_Vector}, \eqref{p1minus} and \eqref{p2minus} have been derived in a revisited work \cite{Lindner:2016bgg}, that differs from the expressions derived in a previous work \cite{Queiroz:2014zfa} used in the reference \cite{Cogollo:2014tra}. The differences between the two results are the signal of the second integral in \eqref{eq:Delta_a_Doubly_Vector} and the signal of the first term of the polynomial \eqref{p1minus}. This is the reason why the contribution of the $U^{++}$ boson has changed of sign when compared with the previous work \cite{Cogollo:2014tra}.

 \subsubsection{Neutral Fermion - Charged Gauge Boson}

In the case where a neutral lepton couples to a charged gauge boson and a muon through the following general lagrangian

\begin{equation} \label{eq:LagrangianChargedGaugeBoson}
  \mathcal{L}_\mathrm{int} = g_{ V }^{ ij }W^{ \prime + }_{ \mu  }\overline { N_{ i } } \gamma ^{ \mu  }\ell _{ j }+g_{ A }^{ ij }W^{ \prime + }_{ \mu  }\overline { N_{ i } } \gamma ^{ \mu  }\gamma ^{ 5 }\ell _{ j } + \textrm{h.c.}
\end{equation}

a contribution to \gmu is generated

\begin{eqnarray}
&&
\Delta a_{ \mu  }(N,W^{ \prime  })=-\frac { m_{ \mu  }^{ 2 } }{ 8\pi ^{ 2 }M_{ W^{ \prime  } }^{ 2 } } \int _{ 0 }^{ 1 } dx\frac { g^{ 2 }_{ V }{ P }_{ 1 }^{ + }(x)+g^{ 2 }_{ A }{ P }_{ 1 }^{ - }(x) }{ { \epsilon  }_{ f }^{ 2 }\lambda ^{ 2 }(1-x)(1-{ \epsilon  }_{ f }^{ -2 }x)+x } ,
\end{eqnarray} 

where $\lambda =  m_{\mu}/M_{W^{\prime}}$, ${ \epsilon  }_{ f }=m_{ { N }_{ f } }/m_{ \mu  }$ and $P_{1}^{\pm}$ is defined in the Eq. \eqref{p1poly}. For the first model, model with doubly charged gauge boson, $m_{ { N }_{ f } }={ m }_{ \nu  }$ and the mediators are the $V$ bosons. As for the second model, model without exotic electric charges, $m_{ { N }_{ f } }= { m }_{ N }={ m }_{ N^{'} }$, and the mediators are the $K^{+}, X^{+}$ bosons. 

\subsubsection{Charged lepton - Neutral Gauge Boson.}

In the case where a charged lepton couples to a neutral gauge boson and a muon through the following general lagrangian

\begin{equation} \label{eq:LagrangianNeutralGaugeBoson}
  \mathcal{L}_\mathrm{int} = g_{ V }^{ ij }Z^{ \prime  }_{ \mu  }\overline { E_{ i } } \gamma ^{ \mu  }\ell _{ j }+g_{ A }^{ ij }Z^{ \prime  }_{ \mu  }\overline { E_{ i } } \gamma ^{ \mu  }\gamma ^{ 5 }\ell _{ j } + \textrm{h.c.}
\end{equation}

a contribution to \gmu is generated

\begin{eqnarray}
&&
\Delta a_{ \mu  }(Z^{ \prime  })=\frac { m_{ \mu  }^{ 2 } }{ 8\pi ^{ 2 }M_{ Z^{ \prime  } }^{ 2 } } \int _{ 0 }^{ 1 } dx\frac { g^{ 2 }_{ V }{ P }_{ 2 }^{ + }(x)+g^{ 2 }_{ A }{ P }_{ 2 }^{ - }(x) }{ (1-x)(1-\lambda ^{ 2 }x)+{ \epsilon  }_{ f }^{ 2 }\lambda ^{ 2 }x } ,
\label{vectormuon1}
\end{eqnarray} 

where $\lambda = m_{\mu}/M_{Z^{\prime}}$, ${ \epsilon  }_{ f }=m_{ E_{f} }/m_{ \mu  }$,  and $P_{2}^{\pm}$ is defined in the Eq. \eqref{p2poly}. For the first model, model with doubly charged gauge boson, ${ m }_{ E_{f} }={ m }_{ \mu  }$ and the mediators are the ${ Z }^{ ' }$ and the ${ Z }_{ N }$ bosons. For the second model, model without exotic electric charge, ${ m }_{ E_{f} }={ m }_{ \mu  }$ and the mediator is the ${ Z }^{ ' }$ boson. As for the third model, model with exotic electric charges,  ${ m }_{ E_{f} }={ m }_{ E }={ m }_{ E' }$ for when the mediators are the bosons $K^{0}, X^{0}$, and ${ m }_{ E_{f} }={ m }_{ \mu  }$ for when the mediator is the ${ Z }^{ ' }$ boson.

\subsubsection{Neutral Scalar}
\label{NS}

If there are additional electrically neutral scalar fields in a model, interacting with muons through the following Lagrangian

\begin{equation}
  \mathcal{L}_\textrm{int} = g_{s}^{ij}  \phi\, \overline{\ell_i}\, \ell_j + i g_{p}^{ij}  \phi\, \overline{\ell_i} \gamma^5 \ell_j,
  \label{Eq:neutralscalar}
\end{equation}

they will induce a correction to \gmu given by:

\begin{eqnarray}
&&
\Delta a_{ \mu  }(\phi )=\frac { m_{ \mu  }^{ 2 } }{ 8\pi ^{ 2 }M_{ \phi  }^{ 2 } } \int _{ 0 }^{ 1 } dx\frac { g^{ 2 }_{ s }{ P }_{ 3 }^{ + }(x)+g^{ 2 }_{ p }{ P }_{ 3 }^{ - }(x) }{ (1-x)(1-\lambda ^{ 2 }x)+{ \epsilon  }_{ f }^{ 2 }\lambda ^{ 2 }x },
\label{DeltaNeutralScalar}
\end{eqnarray} 

where

\begin{equation}
{ P }_{ 3 }^{ \pm  }={ x }^{ 2 }\left( 1-x\pm { \epsilon  }_{ f } \right),
\end{equation}{}

and with $\epsilon_f ={ m }_{ f }/{ m }_{ \mu  }$ and $\lambda =m_{ \mu  }/m_{ \phi  }$. From \eqref{DeltaNeutralScalar} we have derived the contributions to the anomaly of the neutral scalars $\zeta^{0 \star}$ and $\zeta^{\prime 0 \star}$. In both cases ${ m }_{ f }={ m }_{ \mu  }$. As for the coupling constants we have $g_{p}=0$ for $\zeta^{0 \star}$ and $\zeta^{\prime 0 \star}$, $g_{s} \sim \frac{m_{\mu}}{v'}$ for the $\zeta^{0 \star}$ scalar and $g_{s} = \lambda_{2,2} \sim 1$ for the $\zeta^{\prime 0 \star}$ scalar. It is important to emphasize that the contributions to $\Delta a_{\mu}$ are proportional to the square of the coupling constant, then the contribution generated by $\zeta^{\prime 0 \star}$ is $\sim 10^{8}$ times greater than the contribution coming from $\zeta^{0 \star}$.

\subsubsection{Singly Charged Scalar} 

The relevant interaction terms for the contribution to \gmu of a scalar with unit charge, are given by

\begin{equation}
  \mathcal{L}_\textrm{int} = g_{s}^{ij}  \phi^+\, \overline{\nu_i}\, \ell_j + g_{p}^{ij}  \phi^+\, \overline{\nu_i} \gamma^5 \ell_j +\ \mathrm{h.c.}
  \label{eq:singlyscalar}
\end{equation}

In this case, \gmu is given by:

\begin{equation}
    \Delta a_\mu\left(\phi^+\right) = -\frac{1}{8\pi^2} \frac{m_\mu^2}{m_{\phi^+}^2}\int_0^1 \mathrm{d}x\, \sum_f \frac{g_{s}^2 P_4^+(x)+g_{p}^2 P_4^-(x)}{\epsilon_f^2\lambda^2(1-x)\left(1-\epsilon_f^{-2}x\right)+x},\label{eq:Delta_a_Singly_Scalar}
  \end{equation}
  where
  \begin{equation}
    P_4^\pm(x) = x(1-x)\left(x \pm \epsilon_f\right),\quad \epsilon_f\equiv \frac{m_{f}}{m_\mu},\quad \lambda\equiv \frac{m_\mu}{m_{\phi^+}}.\label{eq:P2_def} 
  \end{equation}

For the singly charged scalars of the $\phi_{1}$ multiplet ($\zeta_{1}^{+}, \zeta_{2}^{+}, \zeta_{3}^{+}$), $g_{s}=g_{p} \sim \frac{m_{\mu}}{v'}$, and their contributions to the  anomaly, besides being negatives, are irrelevants. As for the singly charged scalars of the $\phi_{1}^{\prime}$ multiplet ($\zeta_{1}^{\prime +}, \zeta_{2}^{\prime +}, \zeta_{3}^{\prime +}$), $g_{s}=g_{p} = \lambda_{2,2} \sim 1$, and their contributions are $\sim 10^{8}$ times greater than the contributions coming from
$\phi_{1}$. Besides the contributions of the singly charged scalars belonging to the inert scalar $\phi_{1}^{\prime}$ be large and negatives, (see \f \ref{allinert1}), it is the contribution of the $\zeta^{\prime 0 \star}$ the dominant one, by far. Even when we add up the individual contributions of all the particles of the model, this total contribution basically coincides with the individual contribution of the $\zeta^{\prime 0 \star}$ particle, as we can see in \f \ref{allinert2}.

\begin{figure*}[!htp]
    \centering
    \includegraphics[scale=0.3]{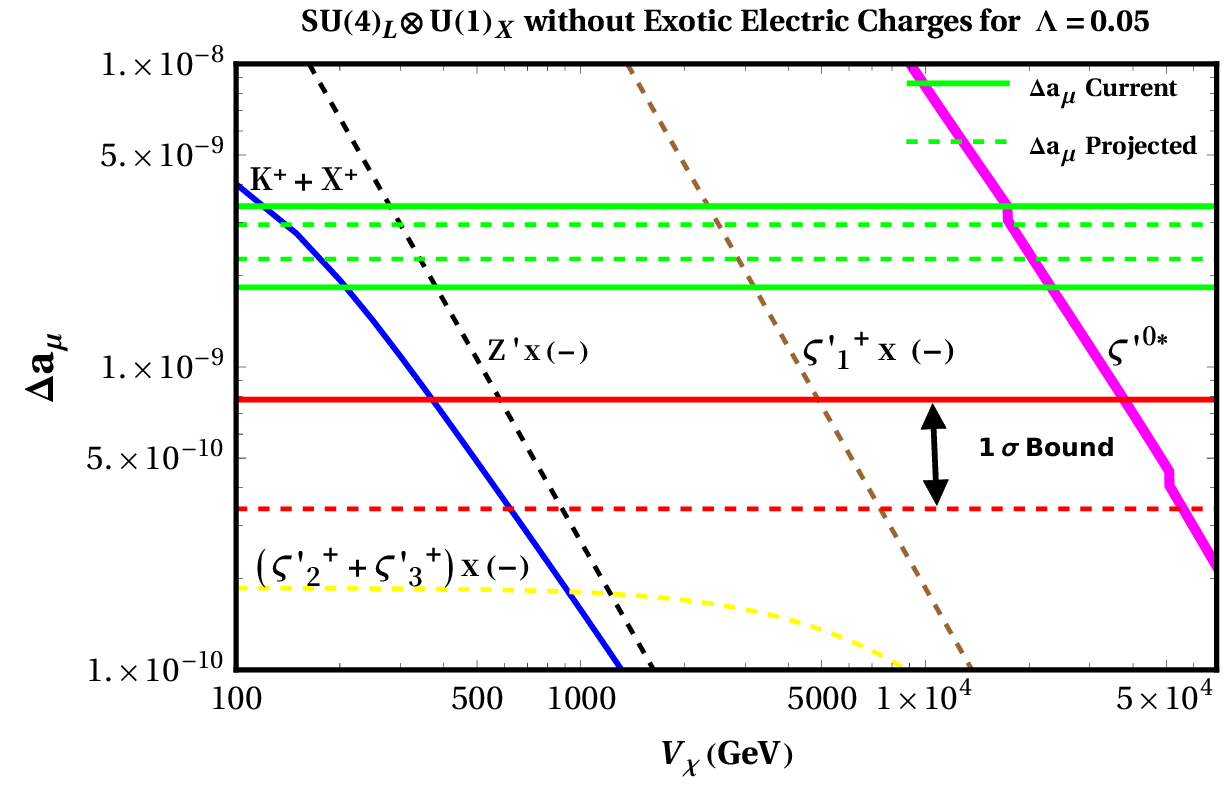}
    \includegraphics[scale=0.3]{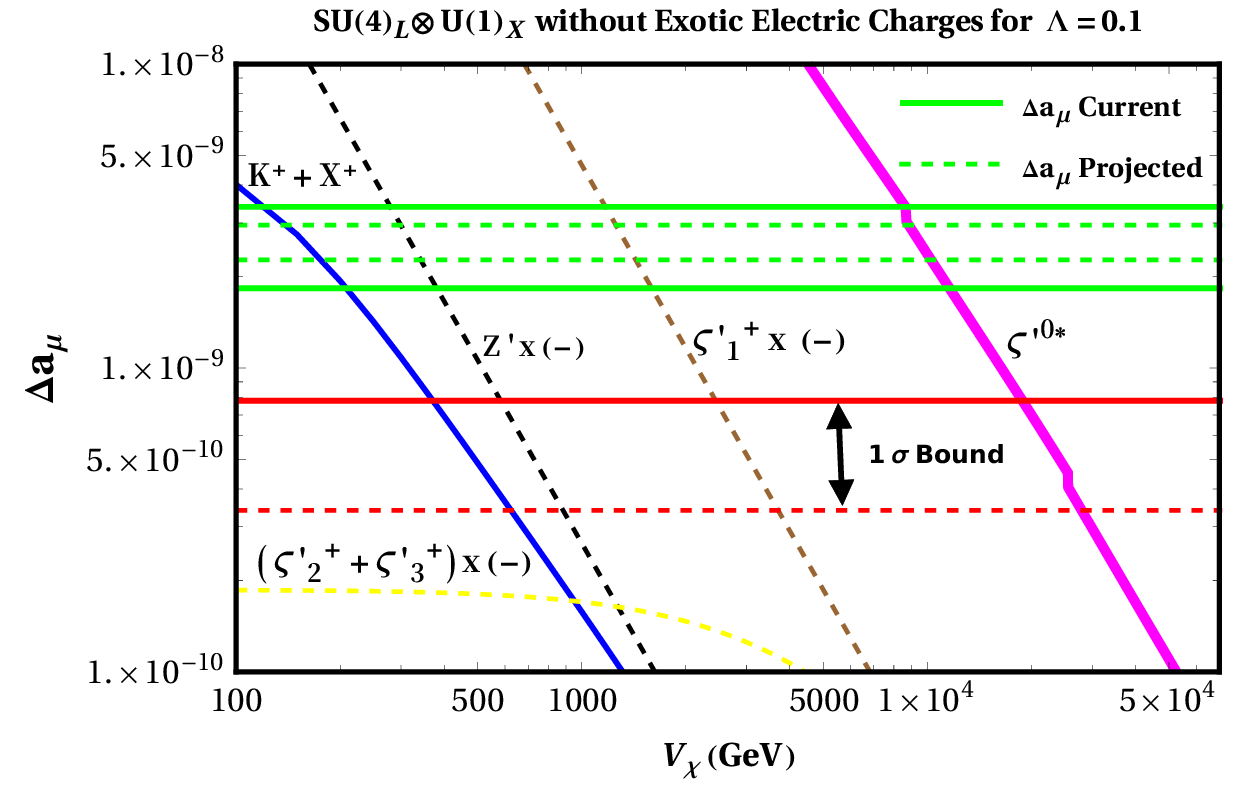}
    \caption{Individual contributions to $\Delta a_{\mu}$ including the inert scalar, as a function of the scale of symmetry breaking $V_\chi$, coming from the model without exotic electric charges, for $M_N$=$M_{N^\prime}=1 TeV$. $\Lambda=0.05$ (up-panel) $\Lambda=0.1$ (down-panel).The current and projected experimental bounds are exhibited with thick and dashed green lines respectively. The 1 $\sigma$ current and projected  error are displayed as thick and dashed red lines respectively.}
    \label{allinert1}
\end{figure*}

\begin{figure*}[!htp]
    \centering
    \includegraphics[scale=0.3]{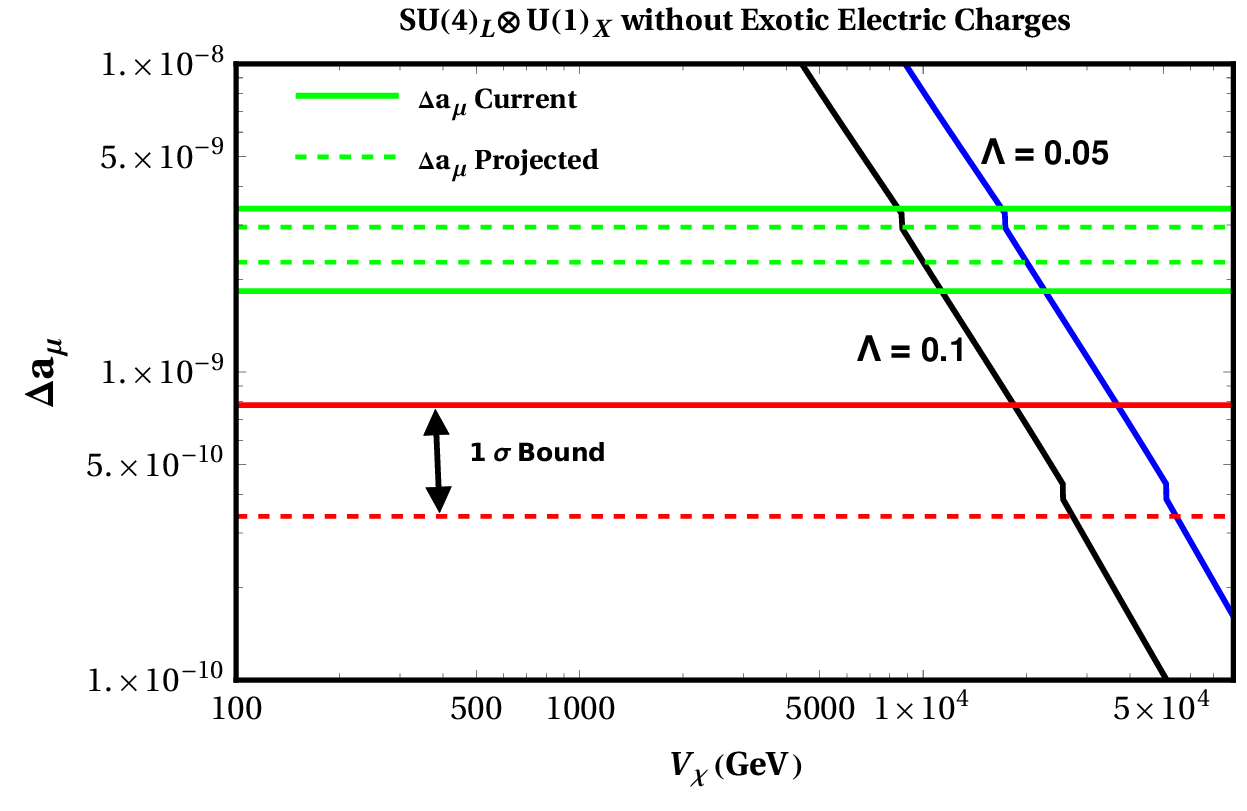}
    \caption{Total contribution to $\Delta a_{\mu}$ including the inert scalar, as a function of the scale of symmetry breaking $V_\chi$, coming from the model without exotic electric charges, for $M_N$=$M_{N^\prime}=1 TeV$. $\Lambda=0.05$ (Blue line) $\Lambda=0.1$ (black line).The current and projected experimental bounds are exhibited with thick and dashed green lines respectively. The 1 $\sigma$ current and projected  error are displayed as thick and dashed red lines respectively.}
    \label{allinert2}
\end{figure*}

In the integral \eqref{eq:Delta_a_Singly_Scalar} we have used    $m_{f}=m_{\nu}$ for the  $\zeta_{1}^{+}$ and $\zeta_{1}^{\prime +}$ particles, and 
$m_{f}=m_{N}=m_{N^{\prime}}$ for $\zeta_{2}^{+}, \zeta_{2}^{\prime +}, \zeta_{3}^{+}, \zeta_{3}^{\prime +}$.

 \end{widetext}

\bibliography{references}

\begin{thebibliography}{77}%
\makeatletter
\providecommand \@ifxundefined [1]{%
 \@ifx{#1\undefined}
}%
\providecommand \@ifnum [1]{%
 \ifnum #1\expandafter \@firstoftwo
 \else \expandafter \@secondoftwo
 \fi
}%
\providecommand \@ifx [1]{%
 \ifx #1\expandafter \@firstoftwo
 \else \expandafter \@secondoftwo
 \fi
}%
\providecommand \natexlab [1]{#1}%
\providecommand \enquote  [1]{``#1''}%
\providecommand \bibnamefont  [1]{#1}%
\providecommand \bibfnamefont [1]{#1}%
\providecommand \citenamefont [1]{#1}%
\providecommand \href@noop [0]{\@secondoftwo}%
\providecommand \href [0]{\begingroup \@sanitize@url \@href}%
\providecommand \@href[1]{\@@startlink{#1}\@@href}%
\providecommand \@@href[1]{\endgroup#1\@@endlink}%
\providecommand \@sanitize@url [0]{\catcode `\\12\catcode `\$12\catcode
  `\&12\catcode `\#12\catcode `\^12\catcode `\_12\catcode `\%12\relax}%
\providecommand \@@startlink[1]{}%
\providecommand \@@endlink[0]{}%
\providecommand \url  [0]{\begingroup\@sanitize@url \@url }%
\providecommand \@url [1]{\endgroup\@href {#1}{\urlprefix }}%
\providecommand \urlprefix  [0]{URL }%
\providecommand \Eprint [0]{\href }%
\providecommand \doibase [0]{http://dx.doi.org/}%
\providecommand \selectlanguage [0]{\@gobble}%
\providecommand \bibinfo  [0]{\@secondoftwo}%
\providecommand \bibfield  [0]{\@secondoftwo}%
\providecommand \translation [1]{[#1]}%
\providecommand \BibitemOpen [0]{}%
\providecommand \bibitemStop [0]{}%
\providecommand \bibitemNoStop [0]{.\EOS\space}%
\providecommand \EOS [0]{\spacefactor3000\relax}%
\providecommand \BibitemShut  [1]{\csname bibitem#1\endcsname}%
\let\auto@bib@innerbib\@empty
\bibitem [{\citenamefont {Pisano}\ and\ \citenamefont
  {Pleitez}(1995)}]{Pisano:1994tf}%
  \BibitemOpen
  \bibfield  {author} {\bibinfo {author} {\bibfnamefont {F.}~\bibnamefont
  {Pisano}}\ and\ \bibinfo {author} {\bibfnamefont {V.}~\bibnamefont
  {Pleitez}},\ }\href {\doibase 10.1103/PhysRevD.51.3865} {\bibfield  {journal}
  {\bibinfo  {journal} {Phys. Rev.}\ }\textbf {\bibinfo {volume} {D51}},\
  \bibinfo {pages} {3865} (\bibinfo {year} {1995})},\ \Eprint
  {http://arxiv.org/abs/hep-ph/9401272} {arXiv:hep-ph/9401272 [hep-ph]}
  \BibitemShut {NoStop}%
\bibitem [{\citenamefont {Cogollo}(2015{\natexlab{a}})}]{Cogollo:2014tra}%
  \BibitemOpen
  \bibfield  {author} {\bibinfo {author} {\bibfnamefont {D.}~\bibnamefont
  {Cogollo}},\ }\href {\doibase 10.1142/S0217751X15500384} {\bibfield
  {journal} {\bibinfo  {journal} {Int. J. Mod. Phys.}\ }\textbf {\bibinfo
  {volume} {A30}},\ \bibinfo {pages} {1550038} (\bibinfo {year}
  {2015}{\natexlab{a}})},\ \Eprint {http://arxiv.org/abs/1409.8115}
  {arXiv:1409.8115 [hep-ph]} \BibitemShut {NoStop}%
\bibitem [{\citenamefont {Cogollo}(2014)}]{Cogollo:2014aka}%
  \BibitemOpen
  \bibfield  {author} {\bibinfo {author} {\bibfnamefont {D.}~\bibnamefont
  {Cogollo}},\ }\href {\doibase 10.3844/pisp.2015.42.50} {\  (\bibinfo {year}
  {2014}),\ 10.3844/pisp.2015.42.50},\ \Eprint {http://arxiv.org/abs/1411.2810}
  {arXiv:1411.2810 [hep-ph]} \BibitemShut {NoStop}%
\bibitem [{\citenamefont {Cogollo}(2015{\natexlab{b}})}]{Cogollo:2015fpa}%
  \BibitemOpen
  \bibfield  {author} {\bibinfo {author} {\bibfnamefont {D.}~\bibnamefont
  {Cogollo}},\ }\href {\doibase 10.1142/S0217751X15501870} {\bibfield
  {journal} {\bibinfo  {journal} {Int. J. Mod. Phys.}\ }\textbf {\bibinfo
  {volume} {A30}},\ \bibinfo {pages} {1550187} (\bibinfo {year}
  {2015}{\natexlab{b}})},\ \Eprint {http://arxiv.org/abs/1508.01492}
  {arXiv:1508.01492 [hep-ph]} \BibitemShut {NoStop}%
\bibitem [{\citenamefont {Lindner}\ \emph {et~al.}(2018)\citenamefont
  {Lindner}, \citenamefont {Platscher},\ and\ \citenamefont
  {Queiroz}}]{Lindner:2016bgg}%
  \BibitemOpen
  \bibfield  {author} {\bibinfo {author} {\bibfnamefont {M.}~\bibnamefont
  {Lindner}}, \bibinfo {author} {\bibfnamefont {M.}~\bibnamefont {Platscher}},
  \ and\ \bibinfo {author} {\bibfnamefont {F.~S.}\ \bibnamefont {Queiroz}},\
  }\href {\doibase 10.1016/j.physrep.2017.12.001} {\bibfield  {journal}
  {\bibinfo  {journal} {Phys. Rept.}\ }\textbf {\bibinfo {volume} {731}},\
  \bibinfo {pages} {1} (\bibinfo {year} {2018})},\ \Eprint
  {http://arxiv.org/abs/1610.06587} {arXiv:1610.06587 [hep-ph]} \BibitemShut
  {NoStop}%
\bibitem [{\citenamefont {Queiroz}\ and\ \citenamefont
  {Shepherd}(2014)}]{Queiroz:2014zfa}%
  \BibitemOpen
  \bibfield  {author} {\bibinfo {author} {\bibfnamefont {F.~S.}\ \bibnamefont
  {Queiroz}}\ and\ \bibinfo {author} {\bibfnamefont {W.}~\bibnamefont
  {Shepherd}},\ }\href {\doibase 10.1103/PhysRevD.89.095024} {\bibfield
  {journal} {\bibinfo  {journal} {Phys. Rev. D}\ }\textbf {\bibinfo {volume}
  {89}},\ \bibinfo {pages} {095024} (\bibinfo {year} {2014})},\ \Eprint
  {http://arxiv.org/abs/1403.2309} {arXiv:1403.2309 [hep-ph]} \BibitemShut
  {NoStop}%
\bibitem [{\citenamefont {Tanabashi}\ \emph {et~al.}(2018)\citenamefont
  {Tanabashi} \emph {et~al.}}]{Tanabashi:2018oca}%
  \BibitemOpen
  \bibfield  {author} {\bibinfo {author} {\bibfnamefont {M.}~\bibnamefont
  {Tanabashi}} \emph {et~al.} (\bibinfo {collaboration} {Particle Data
  Group}),\ }\href {\doibase 10.1103/PhysRevD.98.030001} {\bibfield  {journal}
  {\bibinfo  {journal} {Phys. Rev.}\ }\textbf {\bibinfo {volume} {D98}},\
  \bibinfo {pages} {030001} (\bibinfo {year} {2018})}\BibitemShut {NoStop}%
\bibitem [{\citenamefont {Grange}\ \emph {et~al.}(2015)\citenamefont {Grange}
  \emph {et~al.}}]{Grange:2015fou}%
  \BibitemOpen
  \bibfield  {author} {\bibinfo {author} {\bibfnamefont {J.}~\bibnamefont
  {Grange}} \emph {et~al.} (\bibinfo {collaboration} {Muon g-2}),\ }\href@noop
  {} {\  (\bibinfo {year} {2015})},\ \Eprint {http://arxiv.org/abs/1501.06858}
  {arXiv:1501.06858 [physics.ins-det]} \BibitemShut {NoStop}%
\bibitem [{\citenamefont {Abe}\ \emph {et~al.}(2019)\citenamefont {Abe} \emph
  {et~al.}}]{Abe:2019thb}%
  \BibitemOpen
  \bibfield  {author} {\bibinfo {author} {\bibfnamefont {M.}~\bibnamefont
  {Abe}} \emph {et~al.},\ }\href {\doibase 10.1093/ptep/ptz030} {\bibfield
  {journal} {\bibinfo  {journal} {PTEP}\ }\textbf {\bibinfo {volume} {2019}},\
  \bibinfo {pages} {053C02} (\bibinfo {year} {2019})},\ \Eprint
  {http://arxiv.org/abs/1901.03047} {arXiv:1901.03047 [physics.ins-det]}
  \BibitemShut {NoStop}%
\bibitem [{\citenamefont {Pisano}\ and\ \citenamefont
  {Pleitez}(1992)}]{Pisano:1991ee}%
  \BibitemOpen
  \bibfield  {author} {\bibinfo {author} {\bibfnamefont {F.}~\bibnamefont
  {Pisano}}\ and\ \bibinfo {author} {\bibfnamefont {V.}~\bibnamefont
  {Pleitez}},\ }\href {\doibase 10.1103/PhysRevD.46.410} {\bibfield  {journal}
  {\bibinfo  {journal} {Phys. Rev.}\ }\textbf {\bibinfo {volume} {D46}},\
  \bibinfo {pages} {410} (\bibinfo {year} {1992})},\ \Eprint
  {http://arxiv.org/abs/hep-ph/9206242} {arXiv:hep-ph/9206242 [hep-ph]}
  \BibitemShut {NoStop}%
\bibitem [{\citenamefont {Foot}\ \emph {et~al.}(1993)\citenamefont {Foot},
  \citenamefont {Hernandez}, \citenamefont {Pisano},\ and\ \citenamefont
  {Pleitez}}]{Foot:1992rh}%
  \BibitemOpen
  \bibfield  {author} {\bibinfo {author} {\bibfnamefont {R.}~\bibnamefont
  {Foot}}, \bibinfo {author} {\bibfnamefont {O.~F.}\ \bibnamefont {Hernandez}},
  \bibinfo {author} {\bibfnamefont {F.}~\bibnamefont {Pisano}}, \ and\ \bibinfo
  {author} {\bibfnamefont {V.}~\bibnamefont {Pleitez}},\ }\href {\doibase
  10.1103/PhysRevD.47.4158} {\bibfield  {journal} {\bibinfo  {journal} {Phys.
  Rev.}\ }\textbf {\bibinfo {volume} {D47}},\ \bibinfo {pages} {4158} (\bibinfo
  {year} {1993})},\ \Eprint {http://arxiv.org/abs/hep-ph/9207264}
  {arXiv:hep-ph/9207264 [hep-ph]} \BibitemShut {NoStop}%
\bibitem [{\citenamefont {Fregolente}\ and\ \citenamefont
  {Tonasse}(2003)}]{Fregolente:2002nx}%
  \BibitemOpen
  \bibfield  {author} {\bibinfo {author} {\bibfnamefont {D.}~\bibnamefont
  {Fregolente}}\ and\ \bibinfo {author} {\bibfnamefont {M.~D.}\ \bibnamefont
  {Tonasse}},\ }\href {\doibase 10.1016/S0370-2693(03)00037-6} {\bibfield
  {journal} {\bibinfo  {journal} {Phys. Lett.}\ }\textbf {\bibinfo {volume}
  {B555}},\ \bibinfo {pages} {7} (\bibinfo {year} {2003})},\ \Eprint
  {http://arxiv.org/abs/hep-ph/0209119} {arXiv:hep-ph/0209119 [hep-ph]}
  \BibitemShut {NoStop}%
\bibitem [{\citenamefont {Long}\ and\ \citenamefont
  {Lan}(2003)}]{Hoang:2003vj}%
  \BibitemOpen
  \bibfield  {author} {\bibinfo {author} {\bibfnamefont {H.~N.}\ \bibnamefont
  {Long}}\ and\ \bibinfo {author} {\bibfnamefont {N.~Q.}\ \bibnamefont {Lan}},\
  }\href {\doibase 10.1209/epl/i2003-00267-5} {\bibfield  {journal} {\bibinfo
  {journal} {Europhys. Lett.}\ }\textbf {\bibinfo {volume} {64}},\ \bibinfo
  {pages} {571} (\bibinfo {year} {2003})},\ \Eprint
  {http://arxiv.org/abs/hep-ph/0309038} {arXiv:hep-ph/0309038 [hep-ph]}
  \BibitemShut {NoStop}%
\bibitem [{\citenamefont {de~S.~Pires}\ and\ \citenamefont {Rodrigues~da
  Silva}(2007)}]{deS.Pires:2007gi}%
  \BibitemOpen
  \bibfield  {author} {\bibinfo {author} {\bibfnamefont {C.~A.}\ \bibnamefont
  {de~S.~Pires}}\ and\ \bibinfo {author} {\bibfnamefont {P.~S.}\ \bibnamefont
  {Rodrigues~da Silva}},\ }\href {\doibase 10.1088/1475-7516/2007/12/012}
  {\bibfield  {journal} {\bibinfo  {journal} {JCAP}\ }\textbf {\bibinfo
  {volume} {0712}},\ \bibinfo {pages} {012} (\bibinfo {year} {2007})},\ \Eprint
  {http://arxiv.org/abs/0710.2104} {arXiv:0710.2104 [hep-ph]} \BibitemShut
  {NoStop}%
\bibitem [{\citenamefont {Mizukoshi}\ \emph
  {et~al.}(2011{\natexlab{a}})\citenamefont {Mizukoshi}, \citenamefont
  {de~S.~Pires}, \citenamefont {Queiroz},\ and\ \citenamefont {Rodrigues~da
  Silva}}]{Mizukoshi:2010ky}%
  \BibitemOpen
  \bibfield  {author} {\bibinfo {author} {\bibfnamefont {J.~K.}\ \bibnamefont
  {Mizukoshi}}, \bibinfo {author} {\bibfnamefont {C.~A.}\ \bibnamefont
  {de~S.~Pires}}, \bibinfo {author} {\bibfnamefont {F.~S.}\ \bibnamefont
  {Queiroz}}, \ and\ \bibinfo {author} {\bibfnamefont {P.~S.}\ \bibnamefont
  {Rodrigues~da Silva}},\ }\href {\doibase 10.1103/PhysRevD.83.065024}
  {\bibfield  {journal} {\bibinfo  {journal} {Phys. Rev.}\ }\textbf {\bibinfo
  {volume} {D83}},\ \bibinfo {pages} {065024} (\bibinfo {year}
  {2011}{\natexlab{a}})},\ \Eprint {http://arxiv.org/abs/1010.4097}
  {arXiv:1010.4097 [hep-ph]} \BibitemShut {NoStop}%
\bibitem [{\citenamefont {Profumo}\ and\ \citenamefont
  {Queiroz}(2014)}]{Profumo:2013sca}%
  \BibitemOpen
  \bibfield  {author} {\bibinfo {author} {\bibfnamefont {S.}~\bibnamefont
  {Profumo}}\ and\ \bibinfo {author} {\bibfnamefont {F.~S.}\ \bibnamefont
  {Queiroz}},\ }\href {\doibase 10.1140/epjc/s10052-014-2960-x} {\bibfield
  {journal} {\bibinfo  {journal} {Eur. Phys. J.}\ }\textbf {\bibinfo {volume}
  {C74}},\ \bibinfo {pages} {2960} (\bibinfo {year} {2014})},\ \Eprint
  {http://arxiv.org/abs/1307.7802} {arXiv:1307.7802 [hep-ph]} \BibitemShut
  {NoStop}%
\bibitem [{\citenamefont {Dong}\ \emph
  {et~al.}(2013{\natexlab{a}})\citenamefont {Dong}, \citenamefont {Nguyen},\
  and\ \citenamefont {Soa}}]{Dong:2013ioa}%
  \BibitemOpen
  \bibfield  {author} {\bibinfo {author} {\bibfnamefont {P.~V.}\ \bibnamefont
  {Dong}}, \bibinfo {author} {\bibfnamefont {T.~P.}\ \bibnamefont {Nguyen}}, \
  and\ \bibinfo {author} {\bibfnamefont {D.~V.}\ \bibnamefont {Soa}},\ }\href
  {\doibase 10.1103/PhysRevD.88.095014} {\bibfield  {journal} {\bibinfo
  {journal} {Phys. Rev.}\ }\textbf {\bibinfo {volume} {D88}},\ \bibinfo {pages}
  {095014} (\bibinfo {year} {2013}{\natexlab{a}})},\ \Eprint
  {http://arxiv.org/abs/1308.4097} {arXiv:1308.4097 [hep-ph]} \BibitemShut
  {NoStop}%
\bibitem [{\citenamefont {Dong}\ \emph
  {et~al.}(2013{\natexlab{b}})\citenamefont {Dong}, \citenamefont {Hung},\ and\
  \citenamefont {Tham}}]{Dong:2013wca}%
  \BibitemOpen
  \bibfield  {author} {\bibinfo {author} {\bibfnamefont {P.~V.}\ \bibnamefont
  {Dong}}, \bibinfo {author} {\bibfnamefont {H.~T.}\ \bibnamefont {Hung}}, \
  and\ \bibinfo {author} {\bibfnamefont {T.~D.}\ \bibnamefont {Tham}},\ }\href
  {\doibase 10.1103/PhysRevD.87.115003} {\bibfield  {journal} {\bibinfo
  {journal} {Phys. Rev.}\ }\textbf {\bibinfo {volume} {D87}},\ \bibinfo {pages}
  {115003} (\bibinfo {year} {2013}{\natexlab{b}})},\ \Eprint
  {http://arxiv.org/abs/1305.0369} {arXiv:1305.0369 [hep-ph]} \BibitemShut
  {NoStop}%
\bibitem [{\citenamefont {Queiroz}(2015)}]{Queiroz:2013lca}%
  \BibitemOpen
  \bibfield  {author} {\bibinfo {author} {\bibfnamefont {F.~S.}\ \bibnamefont
  {Queiroz}},\ }\href {\doibase 10.1063/1.4883415} {\bibfield  {journal}
  {\bibinfo  {journal} {AIP Conf. Proc.}\ }\textbf {\bibinfo {volume} {1604}},\
  \bibinfo {pages} {83} (\bibinfo {year} {2015})},\ \Eprint
  {http://arxiv.org/abs/1310.3026} {arXiv:1310.3026 [astro-ph.CO]} \BibitemShut
  {NoStop}%
\bibitem [{\citenamefont {Kelso}\ \emph
  {et~al.}(2014{\natexlab{a}})\citenamefont {Kelso}, \citenamefont
  {de~S.~Pires}, \citenamefont {Profumo}, \citenamefont {Queiroz},\ and\
  \citenamefont {Rodrigues~da Silva}}]{Kelso:2013nwa}%
  \BibitemOpen
  \bibfield  {author} {\bibinfo {author} {\bibfnamefont {C.}~\bibnamefont
  {Kelso}}, \bibinfo {author} {\bibfnamefont {C.~A.}\ \bibnamefont
  {de~S.~Pires}}, \bibinfo {author} {\bibfnamefont {S.}~\bibnamefont
  {Profumo}}, \bibinfo {author} {\bibfnamefont {F.~S.}\ \bibnamefont
  {Queiroz}}, \ and\ \bibinfo {author} {\bibfnamefont {P.~S.}\ \bibnamefont
  {Rodrigues~da Silva}},\ }\href {\doibase 10.1140/epjc/s10052-014-2797-3}
  {\bibfield  {journal} {\bibinfo  {journal} {Eur.\ Phys.\ J.\ C}\ }\textbf
  {\bibinfo {volume} {74}},\ \bibinfo {pages} {2797} (\bibinfo {year}
  {2014}{\natexlab{a}})},\ \Eprint {http://arxiv.org/abs/1308.6630}
  {arXiv:1308.6630 [hep-ph]} \BibitemShut {NoStop}%
\bibitem [{\citenamefont {Cogollo}\ \emph {et~al.}(2014)\citenamefont
  {Cogollo}, \citenamefont {Gonzalez-Morales}, \citenamefont {Queiroz},\ and\
  \citenamefont {Teles}}]{Cogollo:2014jia}%
  \BibitemOpen
  \bibfield  {author} {\bibinfo {author} {\bibfnamefont {D.}~\bibnamefont
  {Cogollo}}, \bibinfo {author} {\bibfnamefont {A.~X.}\ \bibnamefont
  {Gonzalez-Morales}}, \bibinfo {author} {\bibfnamefont {F.~S.}\ \bibnamefont
  {Queiroz}}, \ and\ \bibinfo {author} {\bibfnamefont {P.~R.}\ \bibnamefont
  {Teles}},\ }\href {\doibase 10.1088/1475-7516/2014/11/002} {\bibfield
  {journal} {\bibinfo  {journal} {JCAP}\ }\textbf {\bibinfo {volume} {1411}},\
  \bibinfo {pages} {002} (\bibinfo {year} {2014})},\ \Eprint
  {http://arxiv.org/abs/1402.3271} {arXiv:1402.3271 [hep-ph]} \BibitemShut
  {NoStop}%
\bibitem [{\citenamefont {Dong}\ \emph
  {et~al.}(2014{\natexlab{a}})\citenamefont {Dong}, \citenamefont {Huong},
  \citenamefont {Queiroz},\ and\ \citenamefont {Thuy}}]{Dong:2014wsa}%
  \BibitemOpen
  \bibfield  {author} {\bibinfo {author} {\bibfnamefont {P.~V.}\ \bibnamefont
  {Dong}}, \bibinfo {author} {\bibfnamefont {D.~T.}\ \bibnamefont {Huong}},
  \bibinfo {author} {\bibfnamefont {F.~S.}\ \bibnamefont {Queiroz}}, \ and\
  \bibinfo {author} {\bibfnamefont {N.~T.}\ \bibnamefont {Thuy}},\ }\href
  {\doibase 10.1103/PhysRevD.90.075021} {\bibfield  {journal} {\bibinfo
  {journal} {Phys. Rev.}\ }\textbf {\bibinfo {volume} {D90}},\ \bibinfo {pages}
  {075021} (\bibinfo {year} {2014}{\natexlab{a}})},\ \Eprint
  {http://arxiv.org/abs/1405.2591} {arXiv:1405.2591 [hep-ph]} \BibitemShut
  {NoStop}%
\bibitem [{\citenamefont {Dong}\ \emph
  {et~al.}(2014{\natexlab{b}})\citenamefont {Dong}, \citenamefont {Ngan},\ and\
  \citenamefont {Soa}}]{Dong:2014esa}%
  \BibitemOpen
  \bibfield  {author} {\bibinfo {author} {\bibfnamefont {P.~V.}\ \bibnamefont
  {Dong}}, \bibinfo {author} {\bibfnamefont {N.~T.~K.}\ \bibnamefont {Ngan}}, \
  and\ \bibinfo {author} {\bibfnamefont {D.~V.}\ \bibnamefont {Soa}},\ }\href
  {\doibase 10.1103/PhysRevD.90.075019} {\bibfield  {journal} {\bibinfo
  {journal} {Phys. Rev.}\ }\textbf {\bibinfo {volume} {D90}},\ \bibinfo {pages}
  {075019} (\bibinfo {year} {2014}{\natexlab{b}})},\ \Eprint
  {http://arxiv.org/abs/1407.3839} {arXiv:1407.3839 [hep-ph]} \BibitemShut
  {NoStop}%
\bibitem [{\citenamefont {Mambrini}\ \emph {et~al.}(2016)\citenamefont
  {Mambrini}, \citenamefont {Profumo},\ and\ \citenamefont
  {Queiroz}}]{Mambrini:2015sia}%
  \BibitemOpen
  \bibfield  {author} {\bibinfo {author} {\bibfnamefont {Y.}~\bibnamefont
  {Mambrini}}, \bibinfo {author} {\bibfnamefont {S.}~\bibnamefont {Profumo}}, \
  and\ \bibinfo {author} {\bibfnamefont {F.~S.}\ \bibnamefont {Queiroz}},\
  }\href {\doibase 10.1016/j.physletb.2016.07.076} {\bibfield  {journal}
  {\bibinfo  {journal} {Phys. Lett. B}\ }\textbf {\bibinfo {volume} {760}},\
  \bibinfo {pages} {807} (\bibinfo {year} {2016})},\ \Eprint
  {http://arxiv.org/abs/1508.06635} {arXiv:1508.06635 [hep-ph]} \BibitemShut
  {NoStop}%
\bibitem [{\citenamefont {Alves}\ \emph {et~al.}(2017)\citenamefont {Alves},
  \citenamefont {Arcadi}, \citenamefont {Dong}, \citenamefont {Duarte},
  \citenamefont {Queiroz},\ and\ \citenamefont {Valle}}]{Alves:2016fqe}%
  \BibitemOpen
  \bibfield  {author} {\bibinfo {author} {\bibfnamefont {A.}~\bibnamefont
  {Alves}}, \bibinfo {author} {\bibfnamefont {G.}~\bibnamefont {Arcadi}},
  \bibinfo {author} {\bibfnamefont {P.}~\bibnamefont {Dong}}, \bibinfo {author}
  {\bibfnamefont {L.}~\bibnamefont {Duarte}}, \bibinfo {author} {\bibfnamefont
  {F.~S.}\ \bibnamefont {Queiroz}}, \ and\ \bibinfo {author} {\bibfnamefont
  {J.~W.~F.}\ \bibnamefont {Valle}},\ }\href {\doibase
  10.1016/j.physletb.2017.07.056} {\bibfield  {journal} {\bibinfo  {journal}
  {Phys. Lett. B}\ }\textbf {\bibinfo {volume} {772}},\ \bibinfo {pages} {825}
  (\bibinfo {year} {2017})},\ \Eprint {http://arxiv.org/abs/1612.04383}
  {arXiv:1612.04383 [hep-ph]} \BibitemShut {NoStop}%
\bibitem [{\citenamefont {Carvajal}\ \emph {et~al.}(2017)\citenamefont
  {Carvajal}, \citenamefont {Sánchez-Vega},\ and\ \citenamefont
  {Zapata}}]{Carvajal:2017gjj}%
  \BibitemOpen
  \bibfield  {author} {\bibinfo {author} {\bibfnamefont {C.~D.~R.}\
  \bibnamefont {Carvajal}}, \bibinfo {author} {\bibfnamefont {B.~L.}\
  \bibnamefont {Sánchez-Vega}}, \ and\ \bibinfo {author} {\bibfnamefont
  {O.}~\bibnamefont {Zapata}},\ }\href {\doibase 10.1103/PhysRevD.96.115035}
  {\bibfield  {journal} {\bibinfo  {journal} {Phys. Rev.}\ }\textbf {\bibinfo
  {volume} {D96}},\ \bibinfo {pages} {115035} (\bibinfo {year} {2017})},\
  \Eprint {http://arxiv.org/abs/1704.08340} {arXiv:1704.08340 [hep-ph]}
  \BibitemShut {NoStop}%
\bibitem [{\citenamefont {Dong}\ \emph {et~al.}(2018)\citenamefont {Dong},
  \citenamefont {Huong}, \citenamefont {Queiroz}, \citenamefont {Valle},\ and\
  \citenamefont {Vaquera-Araujo}}]{Dong:2017zxo}%
  \BibitemOpen
  \bibfield  {author} {\bibinfo {author} {\bibfnamefont {P.}~\bibnamefont
  {Dong}}, \bibinfo {author} {\bibfnamefont {D.}~\bibnamefont {Huong}},
  \bibinfo {author} {\bibfnamefont {F.~S.}\ \bibnamefont {Queiroz}}, \bibinfo
  {author} {\bibfnamefont {J.~W.~F.}\ \bibnamefont {Valle}}, \ and\ \bibinfo
  {author} {\bibfnamefont {C.}~\bibnamefont {Vaquera-Araujo}},\ }\href
  {\doibase 10.1007/JHEP04(2018)143} {\bibfield  {journal} {\bibinfo  {journal}
  {JHEP}\ }\textbf {\bibinfo {volume} {04}},\ \bibinfo {pages} {143} (\bibinfo
  {year} {2018})},\ \Eprint {http://arxiv.org/abs/1710.06951} {arXiv:1710.06951
  [hep-ph]} \BibitemShut {NoStop}%
\bibitem [{\citenamefont {Montero}\ \emph {et~al.}(2018)\citenamefont
  {Montero}, \citenamefont {Romero},\ and\ \citenamefont
  {Sánchez-Vega}}]{Montero:2017yvy}%
  \BibitemOpen
  \bibfield  {author} {\bibinfo {author} {\bibfnamefont {J.~C.}\ \bibnamefont
  {Montero}}, \bibinfo {author} {\bibfnamefont {A.}~\bibnamefont {Romero}}, \
  and\ \bibinfo {author} {\bibfnamefont {B.~L.}\ \bibnamefont
  {Sánchez-Vega}},\ }\href {\doibase 10.1103/PhysRevD.97.063015} {\bibfield
  {journal} {\bibinfo  {journal} {Phys. Rev.}\ }\textbf {\bibinfo {volume}
  {D97}},\ \bibinfo {pages} {063015} (\bibinfo {year} {2018})},\ \Eprint
  {http://arxiv.org/abs/1709.04535} {arXiv:1709.04535 [hep-ph]} \BibitemShut
  {NoStop}%
\bibitem [{\citenamefont {Arcadi}\ \emph {et~al.}(2018)\citenamefont {Arcadi},
  \citenamefont {Ferreira}, \citenamefont {Goertz}, \citenamefont {Guzzo},
  \citenamefont {Queiroz},\ and\ \citenamefont {Santos}}]{Arcadi:2017xbo}%
  \BibitemOpen
  \bibfield  {author} {\bibinfo {author} {\bibfnamefont {G.}~\bibnamefont
  {Arcadi}}, \bibinfo {author} {\bibfnamefont {C.}~\bibnamefont {Ferreira}},
  \bibinfo {author} {\bibfnamefont {F.}~\bibnamefont {Goertz}}, \bibinfo
  {author} {\bibfnamefont {M.}~\bibnamefont {Guzzo}}, \bibinfo {author}
  {\bibfnamefont {F.~S.}\ \bibnamefont {Queiroz}}, \ and\ \bibinfo {author}
  {\bibfnamefont {A.}~\bibnamefont {Santos}},\ }\href {\doibase
  10.1103/PhysRevD.97.075022} {\bibfield  {journal} {\bibinfo  {journal}
  {Phys.\ Rev.\ D}\ }\textbf {\bibinfo {volume} {97}},\ \bibinfo {pages}
  {075022} (\bibinfo {year} {2018})},\ \Eprint
  {http://arxiv.org/abs/1712.02373} {arXiv:1712.02373 [hep-ph]} \BibitemShut
  {NoStop}%
\bibitem [{\citenamefont {Huong}\ \emph {et~al.}(2019)\citenamefont {Huong},
  \citenamefont {Dinh}, \citenamefont {Thien},\ and\ \citenamefont
  {Van~Dong}}]{Huong:2019vej}%
  \BibitemOpen
  \bibfield  {author} {\bibinfo {author} {\bibfnamefont {D.~T.}\ \bibnamefont
  {Huong}}, \bibinfo {author} {\bibfnamefont {D.~N.}\ \bibnamefont {Dinh}},
  \bibinfo {author} {\bibfnamefont {L.~D.}\ \bibnamefont {Thien}}, \ and\
  \bibinfo {author} {\bibfnamefont {P.}~\bibnamefont {Van~Dong}},\ }\href
  {\doibase 10.1007/JHEP08(2019)051} {\bibfield  {journal} {\bibinfo  {journal}
  {JHEP}\ }\textbf {\bibinfo {volume} {08}},\ \bibinfo {pages} {051} (\bibinfo
  {year} {2019})},\ \Eprint {http://arxiv.org/abs/1906.05240} {arXiv:1906.05240
  [hep-ph]} \BibitemShut {NoStop}%
\bibitem [{\citenamefont {Montero}\ \emph {et~al.}(2001)\citenamefont
  {Montero}, \citenamefont {de~S.~Pires},\ and\ \citenamefont
  {Pleitez}}]{Montero:2000rh}%
  \BibitemOpen
  \bibfield  {author} {\bibinfo {author} {\bibfnamefont {J.~C.}\ \bibnamefont
  {Montero}}, \bibinfo {author} {\bibfnamefont {C.~A.}\ \bibnamefont
  {de~S.~Pires}}, \ and\ \bibinfo {author} {\bibfnamefont {V.}~\bibnamefont
  {Pleitez}},\ }\href {\doibase 10.1016/S0370-2693(01)00185-X} {\bibfield
  {journal} {\bibinfo  {journal} {Phys. Lett.}\ }\textbf {\bibinfo {volume}
  {B502}},\ \bibinfo {pages} {167} (\bibinfo {year} {2001})},\ \Eprint
  {http://arxiv.org/abs/hep-ph/0011296} {arXiv:hep-ph/0011296 [hep-ph]}
  \BibitemShut {NoStop}%
\bibitem [{\citenamefont {Tully}\ and\ \citenamefont
  {Joshi}(2001)}]{Tully:2000kk}%
  \BibitemOpen
  \bibfield  {author} {\bibinfo {author} {\bibfnamefont {M.~B.}\ \bibnamefont
  {Tully}}\ and\ \bibinfo {author} {\bibfnamefont {G.~C.}\ \bibnamefont
  {Joshi}},\ }\href {\doibase 10.1103/PhysRevD.64.011301} {\bibfield  {journal}
  {\bibinfo  {journal} {Phys. Rev.}\ }\textbf {\bibinfo {volume} {D64}},\
  \bibinfo {pages} {011301} (\bibinfo {year} {2001})},\ \Eprint
  {http://arxiv.org/abs/hep-ph/0011172} {arXiv:hep-ph/0011172 [hep-ph]}
  \BibitemShut {NoStop}%
\bibitem [{\citenamefont {Montero}\ \emph {et~al.}(2002)\citenamefont
  {Montero}, \citenamefont {De~S.~Pires},\ and\ \citenamefont
  {Pleitez}}]{Montero:2001ts}%
  \BibitemOpen
  \bibfield  {author} {\bibinfo {author} {\bibfnamefont {J.~C.}\ \bibnamefont
  {Montero}}, \bibinfo {author} {\bibfnamefont {C.~A.}\ \bibnamefont
  {De~S.~Pires}}, \ and\ \bibinfo {author} {\bibfnamefont {V.}~\bibnamefont
  {Pleitez}},\ }\href {\doibase 10.1103/PhysRevD.65.095001} {\bibfield
  {journal} {\bibinfo  {journal} {Phys. Rev.}\ }\textbf {\bibinfo {volume}
  {D65}},\ \bibinfo {pages} {095001} (\bibinfo {year} {2002})},\ \Eprint
  {http://arxiv.org/abs/hep-ph/0112246} {arXiv:hep-ph/0112246 [hep-ph]}
  \BibitemShut {NoStop}%
\bibitem [{\citenamefont {Cortez}\ and\ \citenamefont
  {Tonasse}(2005)}]{Cortez:2005cp}%
  \BibitemOpen
  \bibfield  {author} {\bibinfo {author} {\bibfnamefont {N.~V.}\ \bibnamefont
  {Cortez}}\ and\ \bibinfo {author} {\bibfnamefont {M.~D.}\ \bibnamefont
  {Tonasse}},\ }\href {\doibase 10.1103/PhysRevD.72.073005} {\bibfield
  {journal} {\bibinfo  {journal} {Phys. Rev.}\ }\textbf {\bibinfo {volume}
  {D72}},\ \bibinfo {pages} {073005} (\bibinfo {year} {2005})},\ \Eprint
  {http://arxiv.org/abs/hep-ph/0510143} {arXiv:hep-ph/0510143 [hep-ph]}
  \BibitemShut {NoStop}%
\bibitem [{\citenamefont {Cogollo}\ \emph {et~al.}(2009)\citenamefont
  {Cogollo}, \citenamefont {Diniz},\ and\ \citenamefont
  {de~S.~Pires}}]{Cogollo:2009yi}%
  \BibitemOpen
  \bibfield  {author} {\bibinfo {author} {\bibfnamefont {D.}~\bibnamefont
  {Cogollo}}, \bibinfo {author} {\bibfnamefont {H.}~\bibnamefont {Diniz}}, \
  and\ \bibinfo {author} {\bibfnamefont {C.~A.}\ \bibnamefont {de~S.~Pires}},\
  }\href {\doibase 10.1016/j.physletb.2009.05.060} {\bibfield  {journal}
  {\bibinfo  {journal} {Phys. Lett.}\ }\textbf {\bibinfo {volume} {B677}},\
  \bibinfo {pages} {338} (\bibinfo {year} {2009})},\ \Eprint
  {http://arxiv.org/abs/0903.0370} {arXiv:0903.0370 [hep-ph]} \BibitemShut
  {NoStop}%
\bibitem [{\citenamefont {Queiroz}\ \emph {et~al.}(2010)\citenamefont
  {Queiroz}, \citenamefont {de~S.Pires},\ and\ \citenamefont
  {da~Silva}}]{Queiroz:2010rj}%
  \BibitemOpen
  \bibfield  {author} {\bibinfo {author} {\bibfnamefont {F.}~\bibnamefont
  {Queiroz}}, \bibinfo {author} {\bibfnamefont {C.}~\bibnamefont {de~S.Pires}},
  \ and\ \bibinfo {author} {\bibfnamefont {P.}~\bibnamefont {da~Silva}},\
  }\href {\doibase 10.1103/PhysRevD.82.065018} {\bibfield  {journal} {\bibinfo
  {journal} {Phys.\ Rev.\ D}\ }\textbf {\bibinfo {volume} {82}},\ \bibinfo
  {pages} {065018} (\bibinfo {year} {2010})},\ \Eprint
  {http://arxiv.org/abs/1003.1270} {arXiv:1003.1270 [hep-ph]} \BibitemShut
  {NoStop}%
\bibitem [{\citenamefont {Cogollo}\ \emph {et~al.}(2010)\citenamefont
  {Cogollo}, \citenamefont {Diniz},\ and\ \citenamefont
  {de~S.~Pires}}]{Cogollo:2010jw}%
  \BibitemOpen
  \bibfield  {author} {\bibinfo {author} {\bibfnamefont {D.}~\bibnamefont
  {Cogollo}}, \bibinfo {author} {\bibfnamefont {H.}~\bibnamefont {Diniz}}, \
  and\ \bibinfo {author} {\bibfnamefont {C.~A.}\ \bibnamefont {de~S.~Pires}},\
  }\href {\doibase 10.1016/j.physletb.2010.03.066} {\bibfield  {journal}
  {\bibinfo  {journal} {Phys. Lett.}\ }\textbf {\bibinfo {volume} {B687}},\
  \bibinfo {pages} {400} (\bibinfo {year} {2010})},\ \Eprint
  {http://arxiv.org/abs/1002.1944} {arXiv:1002.1944 [hep-ph]} \BibitemShut
  {NoStop}%
\bibitem [{\citenamefont {Cogollo}\ \emph {et~al.}(2008)\citenamefont
  {Cogollo}, \citenamefont {Diniz}, \citenamefont {de~S.~Pires},\ and\
  \citenamefont {Rodrigues~da Silva}}]{Cogollo:2008zc}%
  \BibitemOpen
  \bibfield  {author} {\bibinfo {author} {\bibfnamefont {D.}~\bibnamefont
  {Cogollo}}, \bibinfo {author} {\bibfnamefont {H.}~\bibnamefont {Diniz}},
  \bibinfo {author} {\bibfnamefont {C.~A.}\ \bibnamefont {de~S.~Pires}}, \ and\
  \bibinfo {author} {\bibfnamefont {P.~S.}\ \bibnamefont {Rodrigues~da
  Silva}},\ }\href {\doibase 10.1140/epjc/s10052-008-0749-5} {\bibfield
  {journal} {\bibinfo  {journal} {Eur. Phys. J.}\ }\textbf {\bibinfo {volume}
  {C58}},\ \bibinfo {pages} {455} (\bibinfo {year} {2008})},\ \Eprint
  {http://arxiv.org/abs/0806.3087} {arXiv:0806.3087 [hep-ph]} \BibitemShut
  {NoStop}%
\bibitem [{\citenamefont {Okada}\ \emph {et~al.}(2016)\citenamefont {Okada},
  \citenamefont {Okada},\ and\ \citenamefont {Orikasa}}]{Okada:2015bxa}%
  \BibitemOpen
  \bibfield  {author} {\bibinfo {author} {\bibfnamefont {H.}~\bibnamefont
  {Okada}}, \bibinfo {author} {\bibfnamefont {N.}~\bibnamefont {Okada}}, \ and\
  \bibinfo {author} {\bibfnamefont {Y.}~\bibnamefont {Orikasa}},\ }\href
  {\doibase 10.1103/PhysRevD.93.073006} {\bibfield  {journal} {\bibinfo
  {journal} {Phys. Rev.}\ }\textbf {\bibinfo {volume} {D93}},\ \bibinfo {pages}
  {073006} (\bibinfo {year} {2016})},\ \Eprint
  {http://arxiv.org/abs/1504.01204} {arXiv:1504.01204 [hep-ph]} \BibitemShut
  {NoStop}%
\bibitem [{\citenamefont {Vien}\ \emph {et~al.}(2019)\citenamefont {Vien},
  \citenamefont {Long},\ and\ \citenamefont
  {Cárcamo~Hernández}}]{Vien:2018otl}%
  \BibitemOpen
  \bibfield  {author} {\bibinfo {author} {\bibfnamefont {V.~V.}\ \bibnamefont
  {Vien}}, \bibinfo {author} {\bibfnamefont {H.~N.}\ \bibnamefont {Long}}, \
  and\ \bibinfo {author} {\bibfnamefont {A.~E.}\ \bibnamefont
  {Cárcamo~Hernández}},\ }\href {\doibase 10.1142/S0217732319500056}
  {\bibfield  {journal} {\bibinfo  {journal} {Mod. Phys. Lett.}\ }\textbf
  {\bibinfo {volume} {A34}},\ \bibinfo {pages} {1950005} (\bibinfo {year}
  {2019})},\ \Eprint {http://arxiv.org/abs/1812.07263} {arXiv:1812.07263
  [hep-ph]} \BibitemShut {NoStop}%
\bibitem [{\citenamefont {Cárcamo~Hernández}\ \emph
  {et~al.}(2018)\citenamefont {Cárcamo~Hernández}, \citenamefont {Long},\
  and\ \citenamefont {Vien}}]{carcamoHernandez:2018iel}%
  \BibitemOpen
  \bibfield  {author} {\bibinfo {author} {\bibfnamefont {A.~E.}\ \bibnamefont
  {Cárcamo~Hernández}}, \bibinfo {author} {\bibfnamefont {H.~N.}\
  \bibnamefont {Long}}, \ and\ \bibinfo {author} {\bibfnamefont {V.~V.}\
  \bibnamefont {Vien}},\ }\href {\doibase 10.1140/epjc/s10052-018-6284-0}
  {\bibfield  {journal} {\bibinfo  {journal} {Eur. Phys. J.}\ }\textbf
  {\bibinfo {volume} {C78}},\ \bibinfo {pages} {804} (\bibinfo {year}
  {2018})},\ \Eprint {http://arxiv.org/abs/1803.01636} {arXiv:1803.01636
  [hep-ph]} \BibitemShut {NoStop}%
\bibitem [{\citenamefont {Nguyen}\ \emph {et~al.}(2018)\citenamefont {Nguyen},
  \citenamefont {Le}, \citenamefont {Hong},\ and\ \citenamefont
  {Hue}}]{Nguyen:2018rlb}%
  \BibitemOpen
  \bibfield  {author} {\bibinfo {author} {\bibfnamefont {T.~P.}\ \bibnamefont
  {Nguyen}}, \bibinfo {author} {\bibfnamefont {T.~T.}\ \bibnamefont {Le}},
  \bibinfo {author} {\bibfnamefont {T.~T.}\ \bibnamefont {Hong}}, \ and\
  \bibinfo {author} {\bibfnamefont {L.~T.}\ \bibnamefont {Hue}},\ }\href
  {\doibase 10.1103/PhysRevD.97.073003} {\bibfield  {journal} {\bibinfo
  {journal} {Phys. Rev.}\ }\textbf {\bibinfo {volume} {D97}},\ \bibinfo {pages}
  {073003} (\bibinfo {year} {2018})},\ \Eprint
  {http://arxiv.org/abs/1802.00429} {arXiv:1802.00429 [hep-ph]} \BibitemShut
  {NoStop}%
\bibitem [{\citenamefont {de~Sousa~Pires}\ \emph {et~al.}(2019)\citenamefont
  {de~Sousa~Pires}, \citenamefont {Ferreira De~Freitas}, \citenamefont {Shu},
  \citenamefont {Huang},\ and\ \citenamefont {Wagner
  Vasconcelos~Olegário}}]{Pires:2018kaj}%
  \BibitemOpen
  \bibfield  {author} {\bibinfo {author} {\bibfnamefont {C.~A.}\ \bibnamefont
  {de~Sousa~Pires}}, \bibinfo {author} {\bibfnamefont {F.}~\bibnamefont
  {Ferreira De~Freitas}}, \bibinfo {author} {\bibfnamefont {J.}~\bibnamefont
  {Shu}}, \bibinfo {author} {\bibfnamefont {L.}~\bibnamefont {Huang}}, \ and\
  \bibinfo {author} {\bibfnamefont {P.}~\bibnamefont {Wagner
  Vasconcelos~Olegário}},\ }\href {\doibase 10.1016/j.physletb.2019.134827}
  {\bibfield  {journal} {\bibinfo  {journal} {Phys. Lett.}\ }\textbf {\bibinfo
  {volume} {B797}},\ \bibinfo {pages} {134827} (\bibinfo {year} {2019})},\
  \Eprint {http://arxiv.org/abs/1812.10570} {arXiv:1812.10570 [hep-ph]}
  \BibitemShut {NoStop}%
\bibitem [{\citenamefont {Cárcamo~Hernández}\ \emph
  {et~al.}(2019{\natexlab{a}})\citenamefont {Cárcamo~Hernández},
  \citenamefont {Pérez-Julve},\ and\ \citenamefont
  {Hidalgo~Velásquez}}]{CarcamoHernandez:2019iwh}%
  \BibitemOpen
  \bibfield  {author} {\bibinfo {author} {\bibfnamefont {A.~E.}\ \bibnamefont
  {Cárcamo~Hernández}}, \bibinfo {author} {\bibfnamefont {N.~A.}\
  \bibnamefont {Pérez-Julve}}, \ and\ \bibinfo {author} {\bibfnamefont
  {Y.}~\bibnamefont {Hidalgo~Velásquez}},\ }\href {\doibase
  10.1103/PhysRevD.100.095025} {\bibfield  {journal} {\bibinfo  {journal}
  {Phys. Rev.}\ }\textbf {\bibinfo {volume} {D100}},\ \bibinfo {pages} {095025}
  (\bibinfo {year} {2019}{\natexlab{a}})},\ \Eprint
  {http://arxiv.org/abs/1907.13083} {arXiv:1907.13083 [hep-ph]} \BibitemShut
  {NoStop}%
\bibitem [{\citenamefont {Cárcamo~Hernández}\ \emph
  {et~al.}(2019{\natexlab{b}})\citenamefont {Cárcamo~Hernández},
  \citenamefont {Hidalgo~Velásquez},\ and\ \citenamefont
  {Pérez-Julve}}]{CarcamoHernandez:2019vih}%
  \BibitemOpen
  \bibfield  {author} {\bibinfo {author} {\bibfnamefont {A.~E.}\ \bibnamefont
  {Cárcamo~Hernández}}, \bibinfo {author} {\bibfnamefont {Y.}~\bibnamefont
  {Hidalgo~Velásquez}}, \ and\ \bibinfo {author} {\bibfnamefont {N.~A.}\
  \bibnamefont {Pérez-Julve}},\ }\href {\doibase
  10.1140/epjc/s10052-019-7325-z} {\bibfield  {journal} {\bibinfo  {journal}
  {Eur. Phys. J.}\ }\textbf {\bibinfo {volume} {C79}},\ \bibinfo {pages} {828}
  (\bibinfo {year} {2019}{\natexlab{b}})},\ \Eprint
  {http://arxiv.org/abs/1905.02323} {arXiv:1905.02323 [hep-ph]} \BibitemShut
  {NoStop}%
\bibitem [{\citenamefont {Cárcamo~Hernández}\ \emph
  {et~al.}(2020)\citenamefont {Cárcamo~Hernández}, \citenamefont {Hue},
  \citenamefont {Kovalenko},\ and\ \citenamefont
  {Long}}]{CarcamoHernandez:2020pnh}%
  \BibitemOpen
  \bibfield  {author} {\bibinfo {author} {\bibfnamefont {A.~E.}\ \bibnamefont
  {Cárcamo~Hernández}}, \bibinfo {author} {\bibfnamefont {L.~T.}\
  \bibnamefont {Hue}}, \bibinfo {author} {\bibfnamefont {S.}~\bibnamefont
  {Kovalenko}}, \ and\ \bibinfo {author} {\bibfnamefont {H.~N.}\ \bibnamefont
  {Long}},\ }\href@noop {} {\  (\bibinfo {year} {2020})},\ \Eprint
  {http://arxiv.org/abs/2001.01748} {arXiv:2001.01748 [hep-ph]} \BibitemShut
  {NoStop}%
\bibitem [{\citenamefont {Alves}\ \emph {et~al.}(2011)\citenamefont {Alves},
  \citenamefont {Ramirez~Barreto}, \citenamefont {Dias}, \citenamefont
  {de~S.Pires}, \citenamefont {Queiroz},\ and\ \citenamefont {Rodrigues~da
  Silva}}]{Alves:2011kc}%
  \BibitemOpen
  \bibfield  {author} {\bibinfo {author} {\bibfnamefont {A.}~\bibnamefont
  {Alves}}, \bibinfo {author} {\bibfnamefont {E.}~\bibnamefont
  {Ramirez~Barreto}}, \bibinfo {author} {\bibfnamefont {A.}~\bibnamefont
  {Dias}}, \bibinfo {author} {\bibfnamefont {C.}~\bibnamefont {de~S.Pires}},
  \bibinfo {author} {\bibfnamefont {F.}~\bibnamefont {Queiroz}}, \ and\
  \bibinfo {author} {\bibfnamefont {P.}~\bibnamefont {Rodrigues~da Silva}},\
  }\href {\doibase 10.1103/PhysRevD.84.115004} {\bibfield  {journal} {\bibinfo
  {journal} {Phys.\ Rev.\ D}\ }\textbf {\bibinfo {volume} {84}},\ \bibinfo
  {pages} {115004} (\bibinfo {year} {2011})},\ \Eprint
  {http://arxiv.org/abs/1109.0238} {arXiv:1109.0238 [hep-ph]} \BibitemShut
  {NoStop}%
\bibitem [{\citenamefont {Cogollo}\ \emph {et~al.}(2012)\citenamefont
  {Cogollo}, \citenamefont {de~Andrade}, \citenamefont {Queiroz},\ and\
  \citenamefont {Rebello~Teles}}]{Cogollo:2012ek}%
  \BibitemOpen
  \bibfield  {author} {\bibinfo {author} {\bibfnamefont {D.}~\bibnamefont
  {Cogollo}}, \bibinfo {author} {\bibfnamefont {A.}~\bibnamefont {de~Andrade}},
  \bibinfo {author} {\bibfnamefont {F.}~\bibnamefont {Queiroz}}, \ and\
  \bibinfo {author} {\bibfnamefont {P.}~\bibnamefont {Rebello~Teles}},\ }\href
  {\doibase 10.1140/epjc/s10052-012-2029-7} {\bibfield  {journal} {\bibinfo
  {journal} {Eur.\ Phys.\ J.\ C}\ }\textbf {\bibinfo {volume} {72}},\ \bibinfo
  {pages} {2029} (\bibinfo {year} {2012})},\ \Eprint
  {http://arxiv.org/abs/1201.1268} {arXiv:1201.1268 [hep-ph]} \BibitemShut
  {NoStop}%
\bibitem [{\citenamefont {Ruiz-Alvarez}\ \emph {et~al.}(2012)\citenamefont
  {Ruiz-Alvarez}, \citenamefont {de~S.Pires}, \citenamefont {Queiroz},
  \citenamefont {Restrepo},\ and\ \citenamefont {Rodrigues~da
  Silva}}]{Alvares:2012qv}%
  \BibitemOpen
  \bibfield  {author} {\bibinfo {author} {\bibfnamefont {J.}~\bibnamefont
  {Ruiz-Alvarez}}, \bibinfo {author} {\bibfnamefont {C.}~\bibnamefont
  {de~S.Pires}}, \bibinfo {author} {\bibfnamefont {F.~S.}\ \bibnamefont
  {Queiroz}}, \bibinfo {author} {\bibfnamefont {D.}~\bibnamefont {Restrepo}}, \
  and\ \bibinfo {author} {\bibfnamefont {P.}~\bibnamefont {Rodrigues~da
  Silva}},\ }\href {\doibase 10.1103/PhysRevD.86.075011} {\bibfield  {journal}
  {\bibinfo  {journal} {Phys.\ Rev.\ D}\ }\textbf {\bibinfo {volume} {86}},\
  \bibinfo {pages} {075011} (\bibinfo {year} {2012})},\ \Eprint
  {http://arxiv.org/abs/1206.5779} {arXiv:1206.5779 [hep-ph]} \BibitemShut
  {NoStop}%
\bibitem [{\citenamefont {Alves}\ \emph {et~al.}(2013)\citenamefont {Alves},
  \citenamefont {Ramirez~Barreto}, \citenamefont {Dias}, \citenamefont
  {de~S.Pires}, \citenamefont {Queiroz},\ and\ \citenamefont {Rodrigues~da
  Silva}}]{Alves:2012yp}%
  \BibitemOpen
  \bibfield  {author} {\bibinfo {author} {\bibfnamefont {A.}~\bibnamefont
  {Alves}}, \bibinfo {author} {\bibfnamefont {E.}~\bibnamefont
  {Ramirez~Barreto}}, \bibinfo {author} {\bibfnamefont {A.}~\bibnamefont
  {Dias}}, \bibinfo {author} {\bibfnamefont {C.}~\bibnamefont {de~S.Pires}},
  \bibinfo {author} {\bibfnamefont {F.~S.}\ \bibnamefont {Queiroz}}, \ and\
  \bibinfo {author} {\bibfnamefont {P.}~\bibnamefont {Rodrigues~da Silva}},\
  }\href {\doibase 10.1140/epjc/s10052-013-2288-y} {\bibfield  {journal}
  {\bibinfo  {journal} {Eur.\ Phys.\ J.\ C}\ }\textbf {\bibinfo {volume}
  {73}},\ \bibinfo {pages} {2288} (\bibinfo {year} {2013})},\ \Eprint
  {http://arxiv.org/abs/1207.3699} {arXiv:1207.3699 [hep-ph]} \BibitemShut
  {NoStop}%
\bibitem [{\citenamefont {Caetano}\ \emph {et~al.}(2013)\citenamefont
  {Caetano}, \citenamefont {de~S.~Pires}, \citenamefont {Rodrigues~da Silva},
  \citenamefont {Cogollo},\ and\ \citenamefont {Queiroz}}]{Caetano:2013nya}%
  \BibitemOpen
  \bibfield  {author} {\bibinfo {author} {\bibfnamefont {W.}~\bibnamefont
  {Caetano}}, \bibinfo {author} {\bibfnamefont {C.~A.}\ \bibnamefont
  {de~S.~Pires}}, \bibinfo {author} {\bibfnamefont {P.~S.}\ \bibnamefont
  {Rodrigues~da Silva}}, \bibinfo {author} {\bibfnamefont {D.}~\bibnamefont
  {Cogollo}}, \ and\ \bibinfo {author} {\bibfnamefont {F.~S.}\ \bibnamefont
  {Queiroz}},\ }\href {\doibase 10.1140/epjc/s10052-013-2607-3} {\bibfield
  {journal} {\bibinfo  {journal} {Eur.\ Phys.\ J.\ C}\ }\textbf {\bibinfo
  {volume} {73}},\ \bibinfo {pages} {2607} (\bibinfo {year} {2013})},\ \Eprint
  {http://arxiv.org/abs/1305.7246} {arXiv:1305.7246 [hep-ph]} \BibitemShut
  {NoStop}%
\bibitem [{\citenamefont {Queiroz}\ \emph {et~al.}(2016)\citenamefont
  {Queiroz}, \citenamefont {Siqueira},\ and\ \citenamefont
  {Valle}}]{Queiroz:2016gif}%
  \BibitemOpen
  \bibfield  {author} {\bibinfo {author} {\bibfnamefont {F.~S.}\ \bibnamefont
  {Queiroz}}, \bibinfo {author} {\bibfnamefont {C.}~\bibnamefont {Siqueira}}, \
  and\ \bibinfo {author} {\bibfnamefont {J.~W.~F.}\ \bibnamefont {Valle}},\
  }\href {\doibase 10.1016/j.physletb.2016.10.057} {\bibfield  {journal}
  {\bibinfo  {journal} {Phys. Lett. B}\ }\textbf {\bibinfo {volume} {763}},\
  \bibinfo {pages} {269} (\bibinfo {year} {2016})},\ \Eprint
  {http://arxiv.org/abs/1608.07295} {arXiv:1608.07295 [hep-ph]} \BibitemShut
  {NoStop}%
\bibitem [{\citenamefont {Ferreira}\ \emph {et~al.}(2019)\citenamefont
  {Ferreira}, \citenamefont {de~Melo}, \citenamefont {Kovalenko}, \citenamefont
  {Pinheiro},\ and\ \citenamefont {Queiroz}}]{Ferreira:2019qpf}%
  \BibitemOpen
  \bibfield  {author} {\bibinfo {author} {\bibfnamefont {M.~M.}\ \bibnamefont
  {Ferreira}}, \bibinfo {author} {\bibfnamefont {T.~B.}\ \bibnamefont
  {de~Melo}}, \bibinfo {author} {\bibfnamefont {S.}~\bibnamefont {Kovalenko}},
  \bibinfo {author} {\bibfnamefont {P.~R.}\ \bibnamefont {Pinheiro}}, \ and\
  \bibinfo {author} {\bibfnamefont {F.~S.}\ \bibnamefont {Queiroz}},\ }\href
  {\doibase 10.1140/epjc/s10052-019-7422-z} {\bibfield  {journal} {\bibinfo
  {journal} {Eur.\ Phys.\ J.\ C}\ }\textbf {\bibinfo {volume} {79}},\ \bibinfo
  {pages} {955} (\bibinfo {year} {2019})},\ \Eprint
  {http://arxiv.org/abs/1903.07634} {arXiv:1903.07634 [hep-ph]} \BibitemShut
  {NoStop}%
\bibitem [{\citenamefont {Kelso}\ \emph
  {et~al.}(2014{\natexlab{b}})\citenamefont {Kelso}, \citenamefont {Pinheiro},
  \citenamefont {Queiroz},\ and\ \citenamefont {Shepherd}}]{Kelso:2013zfa}%
  \BibitemOpen
  \bibfield  {author} {\bibinfo {author} {\bibfnamefont {C.}~\bibnamefont
  {Kelso}}, \bibinfo {author} {\bibfnamefont {P.}~\bibnamefont {Pinheiro}},
  \bibinfo {author} {\bibfnamefont {F.~S.}\ \bibnamefont {Queiroz}}, \ and\
  \bibinfo {author} {\bibfnamefont {W.}~\bibnamefont {Shepherd}},\ }\href
  {\doibase 10.1140/epjc/s10052-014-2808-4} {\bibfield  {journal} {\bibinfo
  {journal} {Eur.\ Phys.\ J.\ C}\ }\textbf {\bibinfo {volume} {74}},\ \bibinfo
  {pages} {2808} (\bibinfo {year} {2014}{\natexlab{b}})},\ \Eprint
  {http://arxiv.org/abs/1312.0051} {arXiv:1312.0051 [hep-ph]} \BibitemShut
  {NoStop}%
\bibitem [{\citenamefont {Kelso}\ \emph
  {et~al.}(2014{\natexlab{c}})\citenamefont {Kelso}, \citenamefont {Long},
  \citenamefont {Martinez},\ and\ \citenamefont {Queiroz}}]{Kelso:2014qka}%
  \BibitemOpen
  \bibfield  {author} {\bibinfo {author} {\bibfnamefont {C.}~\bibnamefont
  {Kelso}}, \bibinfo {author} {\bibfnamefont {H.}~\bibnamefont {Long}},
  \bibinfo {author} {\bibfnamefont {R.}~\bibnamefont {Martinez}}, \ and\
  \bibinfo {author} {\bibfnamefont {F.~S.}\ \bibnamefont {Queiroz}},\ }\href
  {\doibase 10.1103/PhysRevD.90.113011} {\bibfield  {journal} {\bibinfo
  {journal} {Phys.\ Rev.\ D}\ }\textbf {\bibinfo {volume} {90}},\ \bibinfo
  {pages} {113011} (\bibinfo {year} {2014}{\natexlab{c}})},\ \Eprint
  {http://arxiv.org/abs/1408.6203} {arXiv:1408.6203 [hep-ph]} \BibitemShut
  {NoStop}%
\bibitem [{\citenamefont {Binh}\ \emph
  {et~al.}(2015{\natexlab{a}})\citenamefont {Binh}, \citenamefont {Huong},\
  and\ \citenamefont {Long}}]{Binh:2015cba}%
  \BibitemOpen
  \bibfield  {author} {\bibinfo {author} {\bibfnamefont {D.}~\bibnamefont
  {Binh}}, \bibinfo {author} {\bibfnamefont {D.}~\bibnamefont {Huong}}, \ and\
  \bibinfo {author} {\bibfnamefont {H.}~\bibnamefont {Long}},\ }\href {\doibase
  10.7868/S004445101512007X} {\bibfield  {journal} {\bibinfo  {journal} {Zh.\
  Eksp.\ Teor.\ Fiz.}\ }\textbf {\bibinfo {volume} {148}},\ \bibinfo {pages}
  {1115} (\bibinfo {year} {2015}{\natexlab{a}})},\ \Eprint
  {http://arxiv.org/abs/1504.03510} {arXiv:1504.03510 [hep-ph]} \BibitemShut
  {NoStop}%
\bibitem [{\citenamefont {Binh}\ \emph
  {et~al.}(2015{\natexlab{b}})\citenamefont {Binh}, \citenamefont {Huong},
  \citenamefont {Hue},\ and\ \citenamefont {Long}}]{Binh:2015jfz}%
  \BibitemOpen
  \bibfield  {author} {\bibinfo {author} {\bibfnamefont {D.~T.}\ \bibnamefont
  {Binh}}, \bibinfo {author} {\bibfnamefont {D.~T.}\ \bibnamefont {Huong}},
  \bibinfo {author} {\bibfnamefont {L.~T.}\ \bibnamefont {Hue}}, \ and\
  \bibinfo {author} {\bibfnamefont {H.~N.}\ \bibnamefont {Long}},\ }\href
  {\doibase 10.15625/0868-3166/25/1/4582} {\bibfield  {journal} {\bibinfo
  {journal} {Commun.\ Phys.}\ }\textbf {\bibinfo {volume} {25}},\ \bibinfo
  {pages} {29} (\bibinfo {year} {2015}{\natexlab{b}})}\BibitemShut {NoStop}%
\bibitem [{\citenamefont {De~Conto}\ and\ \citenamefont
  {Pleitez}(2017)}]{DeConto:2016ith}%
  \BibitemOpen
  \bibfield  {author} {\bibinfo {author} {\bibfnamefont {G.}~\bibnamefont
  {De~Conto}}\ and\ \bibinfo {author} {\bibfnamefont {V.}~\bibnamefont
  {Pleitez}},\ }\href {\doibase 10.1007/JHEP05(2017)104} {\bibfield  {journal}
  {\bibinfo  {journal} {JHEP}\ }\textbf {\bibinfo {volume} {05}},\ \bibinfo
  {pages} {104} (\bibinfo {year} {2017})},\ \Eprint
  {http://arxiv.org/abs/1603.09691} {arXiv:1603.09691 [hep-ph]} \BibitemShut
  {NoStop}%
\bibitem [{\citenamefont {Cogollo}(2017)}]{Cogollo:2017foz}%
  \BibitemOpen
  \bibfield  {author} {\bibinfo {author} {\bibfnamefont {D.}~\bibnamefont
  {Cogollo}},\ }\href@noop {} {\  (\bibinfo {year} {2017})},\ \Eprint
  {http://arxiv.org/abs/1706.00397} {arXiv:1706.00397 [hep-ph]} \BibitemShut
  {NoStop}%
\bibitem [{\citenamefont {de~Conto~Santos}(2018)}]{Santos:2018qdx}%
  \BibitemOpen
  \bibfield  {author} {\bibinfo {author} {\bibfnamefont {G.}~\bibnamefont
  {de~Conto~Santos}},\ }\emph {\bibinfo {title} {{Electron and muon anomalous
  magnetic dipole moment in the 3-3-1 model with heavy leptons}}},\ \href@noop
  {} {Ph.D. thesis},\ \bibinfo  {school} {Sao Paulo, IFT} (\bibinfo {year}
  {2018})\BibitemShut {NoStop}%
\bibitem [{\citenamefont {de~Jesus}\ \emph
  {et~al.}(2020{\natexlab{a}})\citenamefont {de~Jesus}, \citenamefont
  {Kovalenko}, \citenamefont {Queiroz}, \citenamefont {Pires},\ and\
  \citenamefont {Villamizar}}]{deJesus:2020ngn}%
  \BibitemOpen
  \bibfield  {author} {\bibinfo {author} {\bibfnamefont {A.~S.}\ \bibnamefont
  {de~Jesus}}, \bibinfo {author} {\bibfnamefont {S.}~\bibnamefont {Kovalenko}},
  \bibinfo {author} {\bibfnamefont {F.~S.}\ \bibnamefont {Queiroz}}, \bibinfo
  {author} {\bibfnamefont {C.~A.~S.}\ \bibnamefont {Pires}}, \ and\ \bibinfo
  {author} {\bibfnamefont {Y.~S.~S.}\ \bibnamefont {Villamizar}},\ }\href@noop
  {} {\  (\bibinfo {year} {2020}{\natexlab{a}})},\ \Eprint
  {http://arxiv.org/abs/2003.06440} {arXiv:2003.06440 [hep-ph]} \BibitemShut
  {NoStop}%
\bibitem [{\citenamefont {Dias}\ \emph {et~al.}(2014)\citenamefont {Dias},
  \citenamefont {Pinheiro}, \citenamefont {de~S.~Pires},\ and\ \citenamefont
  {Rodrigues~da Silva}}]{Dias:2013kma}%
  \BibitemOpen
  \bibfield  {author} {\bibinfo {author} {\bibfnamefont {A.~G.}\ \bibnamefont
  {Dias}}, \bibinfo {author} {\bibfnamefont {P.~R.~D.}\ \bibnamefont
  {Pinheiro}}, \bibinfo {author} {\bibfnamefont {C.~A.}\ \bibnamefont
  {de~S.~Pires}}, \ and\ \bibinfo {author} {\bibfnamefont {P.~S.}\ \bibnamefont
  {Rodrigues~da Silva}},\ }\href {\doibase 10.1016/j.aop.2014.06.021}
  {\bibfield  {journal} {\bibinfo  {journal} {Annals Phys.}\ }\textbf {\bibinfo
  {volume} {349}},\ \bibinfo {pages} {232} (\bibinfo {year} {2014})},\ \Eprint
  {http://arxiv.org/abs/1309.6644} {arXiv:1309.6644 [hep-ph]} \BibitemShut
  {NoStop}%
\bibitem [{\citenamefont {Palcu}(2009{\natexlab{a}})}]{Palcu:2009ks}%
  \BibitemOpen
  \bibfield  {author} {\bibinfo {author} {\bibfnamefont {A.}~\bibnamefont
  {Palcu}},\ }\href {\doibase 10.1142/S0217732309030813} {\bibfield  {journal}
  {\bibinfo  {journal} {Mod. Phys. Lett.}\ }\textbf {\bibinfo {volume} {A24}},\
  \bibinfo {pages} {1247} (\bibinfo {year} {2009}{\natexlab{a}})},\ \Eprint
  {http://arxiv.org/abs/0902.1301} {arXiv:0902.1301 [hep-ph]} \BibitemShut
  {NoStop}%
\bibitem [{\citenamefont {Palcu}(2009{\natexlab{b}})}]{Palcu:2009ky}%
  \BibitemOpen
  \bibfield  {author} {\bibinfo {author} {\bibfnamefont {A.}~\bibnamefont
  {Palcu}},\ }\href {\doibase 10.1142/S0217732309031090} {\bibfield  {journal}
  {\bibinfo  {journal} {Mod. Phys. Lett.}\ }\textbf {\bibinfo {volume} {A24}},\
  \bibinfo {pages} {1731} (\bibinfo {year} {2009}{\natexlab{b}})},\ \Eprint
  {http://arxiv.org/abs/0902.1828} {arXiv:0902.1828 [hep-ph]} \BibitemShut
  {NoStop}%
\bibitem [{\citenamefont {Palcu}(2009{\natexlab{c}})}]{Palcu:2009kb}%
  \BibitemOpen
  \bibfield  {author} {\bibinfo {author} {\bibfnamefont {A.}~\bibnamefont
  {Palcu}},\ }\href {\doibase 10.1142/S0217751X09046011} {\bibfield  {journal}
  {\bibinfo  {journal} {Int. J. Mod. Phys.}\ }\textbf {\bibinfo {volume}
  {A24}},\ \bibinfo {pages} {4923} (\bibinfo {year} {2009}{\natexlab{c}})},\
  \Eprint {http://arxiv.org/abs/0902.3756} {arXiv:0902.3756 [hep-ph]}
  \BibitemShut {NoStop}%
\bibitem [{\citenamefont {Riazuddin}\ and\ \citenamefont
  {Fayyazuddin}(2008)}]{Riazuddin:2008yx}%
  \BibitemOpen
  \bibfield  {author} {\bibinfo {author} {\bibnamefont {Riazuddin}}\ and\
  \bibinfo {author} {\bibnamefont {Fayyazuddin}},\ }\href {\doibase
  10.1140/epjc/s10052-008-0665-8} {\bibfield  {journal} {\bibinfo  {journal}
  {Eur. Phys. J.}\ }\textbf {\bibinfo {volume} {C56}},\ \bibinfo {pages} {389}
  (\bibinfo {year} {2008})},\ \Eprint {http://arxiv.org/abs/0803.4267}
  {arXiv:0803.4267 [hep-ph]} \BibitemShut {NoStop}%
\bibitem [{\citenamefont {Long}\ \emph {et~al.}(2016)\citenamefont {Long},
  \citenamefont {Hue},\ and\ \citenamefont {Loi}}]{PhysRevD.94.015007}%
  \BibitemOpen
  \bibfield  {author} {\bibinfo {author} {\bibfnamefont {H.~N.}\ \bibnamefont
  {Long}}, \bibinfo {author} {\bibfnamefont {L.~T.}\ \bibnamefont {Hue}}, \
  and\ \bibinfo {author} {\bibfnamefont {D.~V.}\ \bibnamefont {Loi}},\ }\href
  {\doibase 10.1103/PhysRevD.94.015007} {\bibfield  {journal} {\bibinfo
  {journal} {Phys. Rev. D}\ }\textbf {\bibinfo {volume} {94}},\ \bibinfo
  {pages} {015007} (\bibinfo {year} {2016})}\BibitemShut {NoStop}%
\bibitem [{\citenamefont {Palcu}(2012)}]{PhysRevD.85.113010}%
  \BibitemOpen
  \bibfield  {author} {\bibinfo {author} {\bibfnamefont {A.}~\bibnamefont
  {Palcu}},\ }\href {\doibase 10.1103/PhysRevD.85.113010} {\bibfield  {journal}
  {\bibinfo  {journal} {Phys. Rev. D}\ }\textbf {\bibinfo {volume} {85}},\
  \bibinfo {pages} {113010} (\bibinfo {year} {2012})},\ \Eprint
  {http://arxiv.org/abs/1111.6262} {arXiv:1111.6262 [hep-ph]} \BibitemShut
  {NoStop}%
\bibitem [{\citenamefont {Ponce}\ \emph {et~al.}(2004)\citenamefont {Ponce},
  \citenamefont {Gutiérrez},\ and\ \citenamefont {Sánchez}}]{Ponce_2004}%
  \BibitemOpen
  \bibfield  {author} {\bibinfo {author} {\bibfnamefont {W.~A.}\ \bibnamefont
  {Ponce}}, \bibinfo {author} {\bibfnamefont {D.~A.}\ \bibnamefont
  {Gutiérrez}}, \ and\ \bibinfo {author} {\bibfnamefont {L.~A.}\ \bibnamefont
  {Sánchez}},\ }\href {\doibase 10.1103/physrevd.69.055007} {\bibfield
  {journal} {\bibinfo  {journal} {Physical Review D}\ }\textbf {\bibinfo
  {volume} {69}} (\bibinfo {year} {2004}),\
  10.1103/physrevd.69.055007}\BibitemShut {NoStop}%
\bibitem [{\citenamefont {S\'anchez}\ \emph {et~al.}(2008)\citenamefont
  {S\'anchez}, \citenamefont {Wills-Toro},\ and\ \citenamefont
  {Zuluaga}}]{PhysRevD.77.035008}%
  \BibitemOpen
  \bibfield  {author} {\bibinfo {author} {\bibfnamefont {L.~A.}\ \bibnamefont
  {S\'anchez}}, \bibinfo {author} {\bibfnamefont {L.~A.}\ \bibnamefont
  {Wills-Toro}}, \ and\ \bibinfo {author} {\bibfnamefont {J.~I.}\ \bibnamefont
  {Zuluaga}},\ }\href {\doibase 10.1103/PhysRevD.77.035008} {\bibfield
  {journal} {\bibinfo  {journal} {Phys. Rev. D}\ }\textbf {\bibinfo {volume}
  {77}},\ \bibinfo {pages} {035008} (\bibinfo {year} {2008})}\BibitemShut
  {NoStop}%
\bibitem [{\citenamefont {Villamizar}\ and\ \citenamefont
  {Cogollo}(2020)}]{Mathematicacode}%
  \BibitemOpen
  \bibfield  {author} {\bibinfo {author} {\bibfnamefont {O.~T. Y.~M.}\
  \bibnamefont {Villamizar}, \bibfnamefont {Yoxara~S}}\ and\ \bibinfo {author}
  {\bibfnamefont {D.}~\bibnamefont {Cogollo}},\ }\href@noop {} {\emph {\bibinfo
  {title} {{Mathematica numerical codes of the Muon Anomalous Magnetic Moment
  to 341 Models}}}} (\bibinfo {year} {Accessed in 2020}),\ \bibinfo {note}
  {\url{https://1drv.ms/u/s!AmXaqAgFnBmhhMtAedFJ4G9PW2ub7w?e=othbcq}}\BibitemShut
  {NoStop}%
\bibitem [{\citenamefont {Nepomuceno}\ and\ \citenamefont
  {Meirose}(2020)}]{Nepomuceno:2019eaz}%
  \BibitemOpen
  \bibfield  {author} {\bibinfo {author} {\bibfnamefont {A.}~\bibnamefont
  {Nepomuceno}}\ and\ \bibinfo {author} {\bibfnamefont {B.}~\bibnamefont
  {Meirose}},\ }\href {\doibase 10.1103/PhysRevD.101.035017} {\bibfield
  {journal} {\bibinfo  {journal} {Phys. Rev.}\ }\textbf {\bibinfo {volume}
  {D101}},\ \bibinfo {pages} {035017} (\bibinfo {year} {2020})},\ \Eprint
  {http://arxiv.org/abs/1911.12783} {arXiv:1911.12783 [hep-ph]} \BibitemShut
  {NoStop}%
\bibitem [{\citenamefont {Mizukoshi}\ \emph
  {et~al.}(2011{\natexlab{b}})\citenamefont {Mizukoshi}, \citenamefont
  {de~S.~Pires}, \citenamefont {Queiroz},\ and\ \citenamefont {Rodrigues~da
  Silva}}]{PhysRevD.83.065024}%
  \BibitemOpen
  \bibfield  {author} {\bibinfo {author} {\bibfnamefont {J.~K.}\ \bibnamefont
  {Mizukoshi}}, \bibinfo {author} {\bibfnamefont {C.~A.}\ \bibnamefont
  {de~S.~Pires}}, \bibinfo {author} {\bibfnamefont {F.~S.}\ \bibnamefont
  {Queiroz}}, \ and\ \bibinfo {author} {\bibfnamefont {P.~S.}\ \bibnamefont
  {Rodrigues~da Silva}},\ }\href {\doibase 10.1103/PhysRevD.83.065024}
  {\bibfield  {journal} {\bibinfo  {journal} {Phys. Rev. D}\ }\textbf {\bibinfo
  {volume} {83}},\ \bibinfo {pages} {065024} (\bibinfo {year}
  {2011}{\natexlab{b}})}\BibitemShut {NoStop}%
\bibitem [{\citenamefont {Cata\~no M.}\ \emph {et~al.}(2012)\citenamefont
  {Cata\~no M.}, \citenamefont {Mart\'{\i}nez},\ and\ \citenamefont
  {Ochoa}}]{PhysRevD.86.073015}%
  \BibitemOpen
  \bibfield  {author} {\bibinfo {author} {\bibfnamefont {E.}~\bibnamefont
  {Cata\~no M.}}, \bibinfo {author} {\bibfnamefont {R.}~\bibnamefont
  {Mart\'{\i}nez}}, \ and\ \bibinfo {author} {\bibfnamefont {F.}~\bibnamefont
  {Ochoa}},\ }\href {\doibase 10.1103/PhysRevD.86.073015} {\bibfield  {journal}
  {\bibinfo  {journal} {Phys. Rev. D}\ }\textbf {\bibinfo {volume} {86}},\
  \bibinfo {pages} {073015} (\bibinfo {year} {2012})}\BibitemShut {NoStop}%
\bibitem [{\citenamefont {Long}(1996{\natexlab{a}})}]{PhysRevD.54.4691}%
  \BibitemOpen
  \bibfield  {author} {\bibinfo {author} {\bibfnamefont {H.~N.}\ \bibnamefont
  {Long}},\ }\href {\doibase 10.1103/PhysRevD.54.4691} {\bibfield  {journal}
  {\bibinfo  {journal} {Phys. Rev. D}\ }\textbf {\bibinfo {volume} {54}},\
  \bibinfo {pages} {4691} (\bibinfo {year} {1996}{\natexlab{a}})}\BibitemShut
  {NoStop}%
\bibitem [{\citenamefont {Long}(1996{\natexlab{b}})}]{PhysRevD.53.437}%
  \BibitemOpen
  \bibfield  {author} {\bibinfo {author} {\bibfnamefont {H.~N.}\ \bibnamefont
  {Long}},\ }\href {\doibase 10.1103/PhysRevD.53.437} {\bibfield  {journal}
  {\bibinfo  {journal} {Phys. Rev. D}\ }\textbf {\bibinfo {volume} {53}},\
  \bibinfo {pages} {437} (\bibinfo {year} {1996}{\natexlab{b}})}\BibitemShut
  {NoStop}%
\bibitem [{\citenamefont {de~Jesus}\ \emph
  {et~al.}(2020{\natexlab{b}})\citenamefont {de~Jesus}, \citenamefont
  {Kovalenko}, \citenamefont {Queiroz}, \citenamefont {Sinha},\ and\
  \citenamefont {Siqueira}}]{jesus2020vectorlike}%
  \BibitemOpen
  \bibfield  {author} {\bibinfo {author} {\bibfnamefont {A.~S.}\ \bibnamefont
  {de~Jesus}}, \bibinfo {author} {\bibfnamefont {S.}~\bibnamefont {Kovalenko}},
  \bibinfo {author} {\bibfnamefont {F.~S.}\ \bibnamefont {Queiroz}}, \bibinfo
  {author} {\bibfnamefont {K.}~\bibnamefont {Sinha}}, \ and\ \bibinfo {author}
  {\bibfnamefont {C.}~\bibnamefont {Siqueira}},\ }\href@noop {} {\enquote
  {\bibinfo {title} {Vector-like leptons and inert scalar triplet: Lepton
  flavor violation, $g-2$ and collider searches},}\ } (\bibinfo {year}
  {2020}{\natexlab{b}}),\ \Eprint {http://arxiv.org/abs/2004.01200}
  {arXiv:2004.01200 [hep-ph]} \BibitemShut {NoStop}%
\end{thebibliography}%
\end{document}